\documentclass[pdflatex,sn-mathphys-ay]{sn-jnl}

\usepackage{graphicx}%
\usepackage{multirow}%
\usepackage{amsmath,amssymb,amsfonts}%
\usepackage{amsthm}%
\usepackage[title]{appendix}%
\usepackage{xcolor}%
\usepackage{textcomp}%
\usepackage{manyfoot}%
\usepackage{booktabs}%
\usepackage{algorithm}%
\usepackage{algorithmicx}%
\usepackage{algpseudocode}%
\usepackage{listings}%
\usepackage{siunitx}%
\usepackage{longtable}%
\usepackage{adjustbox}%
\usepackage{array}%
\usepackage{url}%
\hypersetup{hypertexnames=false}%
\renewcommand{\thetable}{\Roman{table}}

\theoremstyle{thmstyleone}%

\theoremstyle{thmstyletwo}%

\theoremstyle{thmstylethree}%

\AtBeginDocument{\raggedbottom}
\providecommand{\linenumbers}{}

\begin{document}

\title[A Blockchain-as-a-Service Solution for TAFES-Compliant Verification]{A Blockchain-as-a-Service Solution for TAFES-Compliant Verification of Fair Trade Certifications}

% Author information is anonymised in the supplied manuscript. Replace the placeholder below with final author metadata before submission.
\author*[1]{\fnm{Nadia} \sur{Dahmani}}\email{nadia.Dahmani@zu.ac.ae}

\author[2]{\fnm{Peihao} \sur{Li}}\email{peihao.li@kaust.edu.sa}

\author[1]{\fnm{Ravishankar} \sur{Sharma}}\email{ravishankar.Sharma@zu.ac.ae}

\affil*[1]{\orgdiv{Information Systems and Technology Management Department},
\orgname{College of Technological Innovation, Zayed University},
\orgaddress{\city{Abu Dhabi}, \country{United Arab Emirates}}}

\affil[2]{\orgdiv{Computer, Electrical and Mathematical Sciences and Engineering Division},
\orgname{King Abdullah University of Science and Technology},
\orgaddress{\city{Thuwal}, \country{Saudi Arabia}}}

\abstract{\textbf{Purpose:} This study addresses the lack of trust in ethical product labels by designing a blockchain platform grounded in the TAFES principles (Transparency, Accountability, Fairness, Ethics, Safety). It aims to bridge the gap between blockchain's theoretical transparency and a responsible, real-world implementation for certification ecosystems.

\textbf{Design/Methodology/Approach:} Using Action Design Research, we developed a proof-of-concept platform for label authentication. A hybrid architecture records critical events on an Ethereum Layer-2 network for security, while supporting evidence is stored off-chain via IPFS and linked via content identifiers. The solution was validated through a coffee supply chain scenario.

\textbf{Findings:} The proof of concept demonstrates how a TAFES-aligned blockchain platform can support verification of label claims without requiring trust in a single intermediary by creating tamper-evident provenance records and auditable certification evidence across multiple stakeholders. The design supports low-cost, near-real-time anchoring of supply chain events while mitigating adoption barriers related to scalability, privacy, and operational viability.

\textbf{Originality/Value:} This research contributes an integrated ethical and technical blueprint for trustworthy label authentication systems by translating TAFES into implementable design requirements and evaluation checks, and validating them through an ADR driven proof of concept. It advances prior work by moving from the question of whether blockchain can help to the question of how it should be implemented responsibly in multi stakeholder certification ecosystems.}


\maketitle

\linenumbers

\section{Introduction}
\label{sec:introduction}

\noindent Ethical and sustainability labels have become mainstream signals in global markets, and empirical work shows many consumers are willing to pay premiums for certified products such as Fairtrade coffee \citep{abdu2021wtpcoffee,consumerWTPsustainableCoffee2024}. Yet these claims are largely \textit{credence attributes}: buyers cannot verify the underlying practices at the point of sale and must rely on assurance mechanisms to reduce information asymmetry \citep{foodCredenceAttributes2023}. As label ecosystems proliferate, consumers face growing choice complexity and may experience confusion and scepticism about what labels actually guarantee \citep{langer2007ecolabelconfusion,sodamin2022fair,lou2024denim}.

\noindent This tension is especially visible in fair trade and adjacent ethical label ecosystems. Traditional certification models typically rely on trusted third parties, periodic audits, and document centric reporting \citep{hilten2020blockchain,katsikouli2020benefits,bernards2022veil}. In globalised multi tier supply chains, evidence is fragmented across organisations and information systems, which limits end to end traceability, reduces timely visibility, and increases the opportunity space for mislabelling, fraud, and greenwashing \citep{balzarova2020conundrum,bernards2022veil,friedman2022sustainability,fani2025cultivatingTrust}. Certification costs and administrative overhead can also disproportionately burden small producers, contributing to ongoing scepticism about whether price premiums translate into fair producer outcomes \citep{bager2022not,kshetri2021developing}.

\noindent Blockchain technology is frequently proposed as a remedy because immutable logs and shared validation can strengthen traceability and auditability \citep{kouhizadeh2021adoption,kshetri2021developing,park2021effect,alt2025dlt}. However, there remains a practical and normative gap between high level claims that ``blockchain increases transparency'' and implementable platform designs that work under real constraints (privacy law, uneven digital capability, interoperability, operational burden) while remaining aligned with fair trade ethics \citep{bager2022not,balzarova2020conundrum,bernards2022veil,bons2020potential,ostern2020blockchainIS}. Put differently, the core problem is not only technical feasibility, but also \textit{responsible operationalisation}: how a platform should be designed so that transparency becomes usable and governance remains fair and safe. This socio technical framing is consistent with governance oriented and review studies that emphasise market design choices, process modelling, and cross stakeholder coordination as determinants of realised value \citep{beck2018governance,frizzobarker2020disruptive,casino2019systematic,mendling2018bpm,bendig2025multisidedPlatforms,feulner2025intermediaryRoles}.

\noindent Recent work in \textit{Electronic Markets} sharpens this problem in three ways. First, empirical supply-chain evidence from the Italian wine industry links blockchain-based traceability to trust creation while showing that digital capability gaps, IoT integration, and intermediary support remain important adoption conditions \citep{fani2025cultivatingTrust}. Second, supply-chain automation research shows that blockchain systems must balance cooperative data sharing with competitive confidentiality, especially where shared records expose sensitive commercial information \citep{lautenschlager2025strikingBalance}. Third, enterprise interoperability research shows that blockchain solutions must interoperate not only with other ledgers but also with legacy organisational systems, data structures, and compliance requirements \citep{mafike2026interoperability}.

\noindent This paper addresses that gap by designing and prototyping a blockchain based label authentication platform that operationalises the TAFES principles (Transparency, Accountability, Fairness, Ethics, and Safety) through a hybrid Ethereum Layer 2 and off chain evidence architecture. The work is conducted as design science, using Action Design Research (ADR) to iteratively build and evaluate a proof of concept artefact in a realistic coffee supply chain scenario \citep{sein2011adr,jensen2019dsr,hevner2004dsr,gregor2013dsr}.

\noindent The study is guided by the following research questions:
\begin{description}
  \item[\textbf{RQ1}] How can a blockchain based platform provide verifiable, consumer facing authentication of ethical labels while maintaining privacy, feasibility, and multi stakeholder usability?
  \item[\textbf{RQ2}] How can TAFES be translated into concrete platform requirements and evaluation checks that address known adoption barriers (e.g., oracle and data quality, interoperability, operational burden)?
  \item[\textbf{RQ3}] How can a hybrid Layer 2 plus off chain evidence design reduce cost and latency sufficiently to support near real time provenance recording without undermining producer fairness?
\end{description}

\noindent The paper makes three contributions. First, it synthesises recurring integrity and coordination weaknesses in traditional certification platforms to motivate explicit design requirements. Second, it operationalises TAFES as actionable principles with implementable requirements and evaluation checks. Third, it demonstrates a proof of concept architecture and an ADR based development logic that can be reused in other label ecosystems.

\noindent The remainder of the paper is structured as follows. Section~\ref{sec:blockchain} reviews certification integrity challenges and the affordances and limits of blockchain for label authentication. Section~\ref{sec:design-adr} introduces TAFES and the ADR approach and shows how principles are translated into requirements and a proof of concept instantiation. Subsequent sections present implementation details, evaluation results, discussion, and conclusions.

% ==========================================================
\section{Blockchain technology and certification integrity}
\label{sec:blockchain}

\subsection{Certification integrity challenges in ethical label ecosystems}
\noindent Across the literature, several recurring weaknesses appear in traditional fair trade certification and related ethical label systems. These weaknesses reduce verification quality, hinder coordination, and undermine consumer trust in label claims \citep{hilten2020blockchain,katsikouli2020benefits,bernards2022veil,zavolokina2020buyersLemons,bauer2020trustedCarData}. Table~\ref{tab-I} summarises key issues and their consequences.

\begin{table}[ht]
\centering
\caption{Key issues in traditional fair trade certification platforms}
\label{tab-I}
\setlength{\tabcolsep}{2pt}\scriptsize\renewcommand{\arraystretch}{0.98}
\begin{tabular}{@{}p{0.17\textwidth}p{0.32\textwidth}p{0.24\textwidth}>{\raggedright\arraybackslash\scriptsize}p{0.21\textwidth}@{}}
\toprule
\textbf{Issue} & \textbf{Description} & \textbf{Impact} & \textbf{Sources} \\
\midrule
Lack of trust among stakeholders &
Multi tier supply chains and complex processing chains limit access to timely, shared information &
Lower cooperation; harder verification and dispute resolution &
\citep{hilten2020blockchain,katsikouli2020benefits,fani2025cultivatingTrust}
\\
\addlinespace
Efficiency tracking limitations &
Production, inventory, logistics and sales are difficult to track without shared digital infrastructure &
Operational inefficiency and higher coordination cost &
\citep{kshetri2021developing,rejeb2020food,lautenschlager2025strikingBalance}
\\
\addlinespace
Limited transparency &
Large networks and siloed systems reduce visibility of operations and evidence across tiers &
Reduced accountability and consumer trust &
\citep{balzarova2020conundrum,bernards2022veil,zavolokina2020buyersLemons,bauer2020trustedCarData}
\\
\addlinespace
Low digital integration &
No common platform connects producers, certifiers, intermediaries, and retailers &
Slow payments; inconsistent records; weak controls &
\citep{erol2021scrutinizing,kouhizadeh2021adoption,fani2025cultivatingTrust,mafike2026interoperability}
\\
\addlinespace
Limited end to end visibility &
Partial lifecycle visibility makes it hard to verify claims across the full chain of custody &
Logistics disputes; limited provenance at point of sale &
\citep{stopfer2024wood,santos2021thirdparty,alt2025dlt}
\\
\addlinespace
Contract and evidence management challenges &
Evidence and contracts are document centric and fragmented across organisations &
Compliance friction and higher risk of tampering or loss &
\citep{agrawal2021textile,nikolakis2018eve,lautenschlager2025strikingBalance}
\\
\bottomrule
\end{tabular}
\end{table}

\noindent The table indicates a consistent pattern. Evidence is fragmented, verification is periodic and difficult to audit across the full chain of custody, and the lack of shared digital infrastructure increases both cost and integrity risk. These problems motivate platform designs that improve provenance visibility and evidence integrity while remaining usable for stakeholders with uneven digital resources \citep{kshetri2021developing,kouhizadeh2021adoption,fani2025cultivatingTrust}. Recent \textit{Electronic Markets} studies echo this diagnosis: wine supply-chain adopters report trust deficits, fragmentation and digital capability gaps \citep{fani2025cultivatingTrust}, while construction supply-chain automation research shows that data transparency must be deliberately balanced against confidentiality in coopetitive networks \citep{lautenschlager2025strikingBalance}.

\subsection{Blockchain affordances for label authentication and provenance}
\begin{figure}[ht]
 \centering
 \makebox[\textwidth][c]{\includegraphics[width=0.95\textwidth]{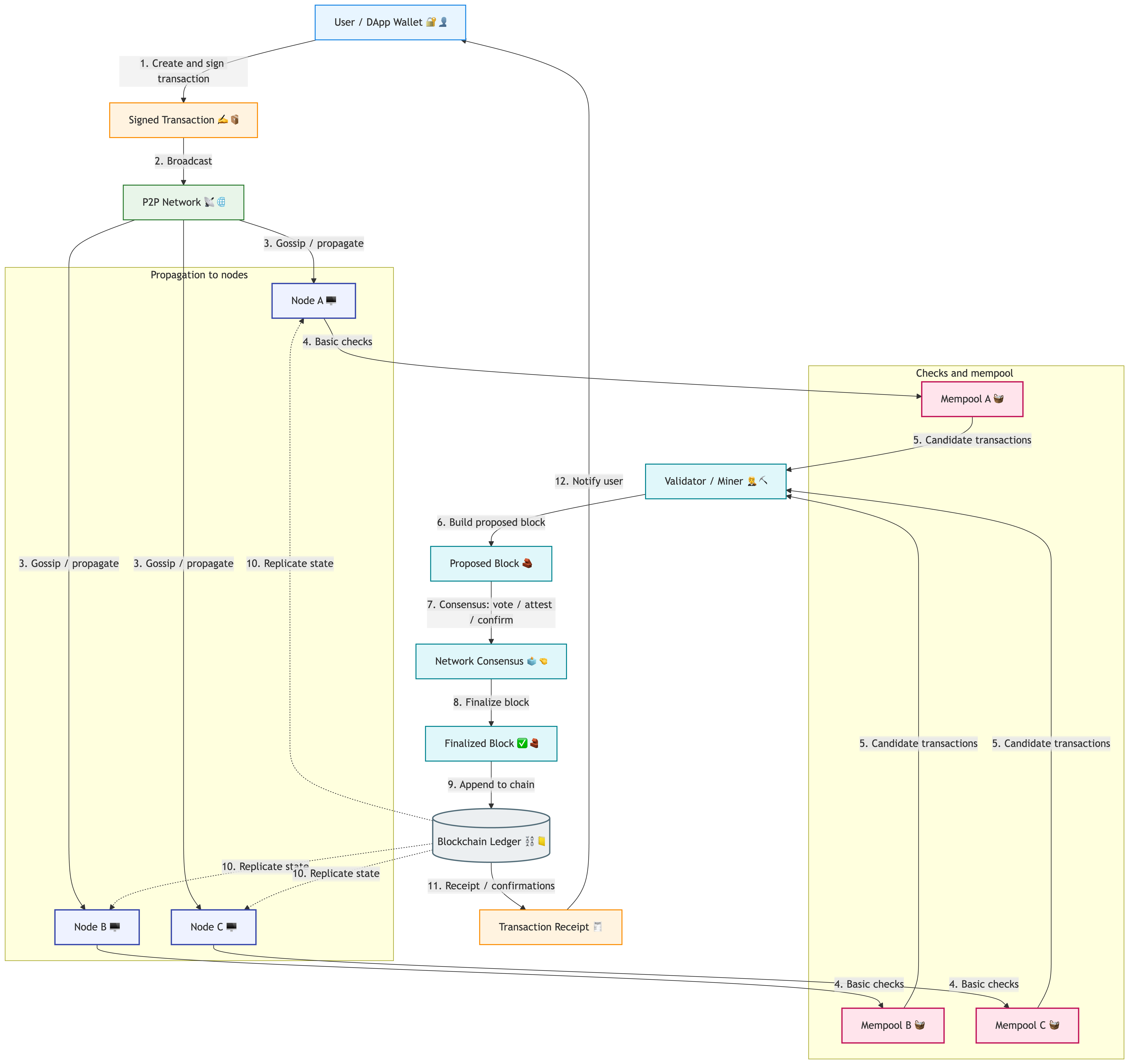}}%
\caption{Blockchain transaction flow (conceptual): transactions are signed and broadcast, validated by multiple nodes, ordered via consensus, finalised into blocks, and replicated as an auditable ledger state across participants.}
 \label{fig:tx-process}
\end{figure}

\noindent Blockchain systems provide decentralised validation, tamper evident logging, and shared auditability, which align with the needs of label authentication and multi stakeholder provenance \citep{guo2020fashion,nikolakis2018eve,santos2021thirdparty,alt2025dlt,bons2020potential}. Figure~\ref{fig:tx-process} illustrates the high level transaction flow that yields the properties most relevant to certification: shared validation (consensus), tamper evidence (immutability), and auditable record replication (transparency).

\noindent Prior studies emphasise four blockchain features that are particularly relevant for ethical label contexts. First, \textit{immutable record keeping} uses cryptographic hashing and distributed agreement so that once evidence pointers or certification states are recorded, later tampering becomes detectable \citep{agrawal2021textile,chandan2023sdgs,hasan2024smartagriculture,alt2025dlt}. Second, \textit{end to end traceability} enables an auditable chain of custody from origin to consumer when lifecycle events are consistently recorded \citep{nikolakis2018eve,park2021effect,stopfer2024wood,fani2025cultivatingTrust}. Third, \textit{smart contracts} allow workflow gates and conditional actions (e.g., certification checks, conditional payments) to execute automatically once verifiable conditions are met \citep{santos2021thirdparty,chandan2023sdgs,mendling2018bpm,lautenschlager2025strikingBalance}. Fourth, \textit{decentralised verification} reduces single point failures and can strengthen cross organisation auditability when stakeholders do not fully trust each other \citep{balzarova2020conundrum,bernards2022veil,friedman2022sustainability,bauer2020trustedCarData,zavolokina2020buyersLemons}.

\begin{table}[ht]
\centering
\caption{Blockchain advantages in fair trade transactions}
\label{tab-bc-affordances}
\setlength{\tabcolsep}{2pt}\scriptsize\renewcommand{\arraystretch}{0.98}
\begin{tabular}{@{}p{0.20\textwidth}p{0.42\textwidth}p{0.30\textwidth}@{}}
\toprule
\textbf{Advantage} & \textbf{Description} & \textbf{Indicative evidence / sources} \\
\midrule
Digital contract and evidence referencing &
Transactions and evidence references can be logged using unique hashes, creating tamper evident audit trails &
Persistent auditing and integrity checks \citep{santos2021thirdparty,agrawal2021textile,alt2020blockchainMarkets} \\
\addlinespace
Visibility and transparency &
Shared ledgers enable consistent provenance views for authorised stakeholders &
Reduced information asymmetry \citep{guo2020fashion,nikolakis2018eve,bauer2020trustedCarData,zavolokina2020buyersLemons} \\
\addlinespace
Stakeholder trust through auditability &
Records are difficult to alter after validation, supporting cross party verification &
Reported trust improvements \citep{kouhizadeh2021adoption,kshetri2021developing,fani2025cultivatingTrust} \\
\addlinespace
Automation via smart contracts &
Rules can be enforced programmatically (e.g., compliance gates, workflow steps, conditional payments) &
Reduced manual overhead \citep{chandan2023sdgs,liu2023impact,mendling2018bpm,lautenschlager2025strikingBalance} \\
\addlinespace
Reduced fraud and tampering opportunities &
Tampering requires overcoming distributed validation, and replicated logs support detection &
Fraud reductions reported \citep{xiaoyong2024certifying,liu2023impact,fani2025cultivatingTrust} \\
\bottomrule
\end{tabular}
\end{table}

\subsection{Adoption challenges and design implications}
\noindent Despite these affordances, using blockchain to address the certification and trust problems summarised in Table~\ref{tab-I} raises well documented challenges \citep{hilten2020blockchain,katsikouli2020benefits,kshetri2021developing,balzarova2020conundrum,bons2020potential,alt2025dlt}. The most relevant challenges for label authentication platforms are as follows.

\begin{enumerate}
\item \textbf{\textit{Communicable verification and limited consumer understanding}} \\
Even when provenance is verifiable, trust gains depend on whether non expert users can interpret what is shown at the point of purchase. A substantial share of consumers report limited understanding of blockchain concepts, which can blunt the value of blockchain backed claims \citep{sodamin2022fair,liu2023impact,contini2023credence,ma2025consumerQuality}.

\item \textbf{\textit{Scalability and cost constraints}} \\
Base layer public blockchains can be costly and throughput limited for high frequency event logging. Many systems therefore use selective anchoring and more scalable layers and data designs \citep{sanka2021scalability,kostamis2021ethereumdatastores}. Review work also highlights scalability and systems integration issues as persistent open problems in blockchain deployments \citep{casino2019systematic,bons2020potential,alt2020blockchainMarkets}.

\item \textbf{\textit{Data quality at entry (oracle problem)}} \\
Blockchains protect data after recording, but do not guarantee that inputs are accurate. Certification settings therefore require governance, audits, and sensor or oracle controls to avoid ``garbage in, garbage out'' \citep{katsikouli2020benefits,caldarelli2020oracle,hassan2023oracle}.

\item \textbf{\textit{Interoperability gaps}} \\
Incompatible ledgers and fragmented digital systems can recreate silos and limit end to end provenance visibility, undermining the shared platform objective \citep{rejeb2020food,kouhizadeh2021adoption,erol2021scrutinizing,mafike2026interoperability}. \textit{Electronic Markets} research further emphasises that interoperability must include blockchain-to-legacy-system integration, organisational scope, data, compliance, and regulatory frameworks rather than only cross-chain asset exchange \citep{mafike2026interoperability}.

\item \textbf{\textit{Privacy and regulatory constraints}} \\
Immutability can conflict with deletion duties and privacy by design expectations. Practical designs must minimise sensitive on chain data while preserving auditability \citep{kshetri2021developing,balzarova2020conundrum,bernards2022veil,lautenschlager2025strikingBalance}.

\item \textbf{\textit{Operational burden and governance}} \\
Key management, onboarding, access control, and operational monitoring can increase adoption friction. Effective systems often combine on chain integrity with off chain governance and service layers \citep{cao2021governance,hilten2020blockchain,beck2018governance,feulner2025intermediaryRoles}. Empirical and theory driven studies also point to readiness, stakeholder pressure, and capability asymmetries as adoption antecedents that must be handled explicitly in platform design \citep{nayal2023antecedents,khan2025greendata,fani2025cultivatingTrust,bendig2025multisidedPlatforms}.
\end{enumerate}

\noindent \textit{Design implications.} These challenges motivate a design posture in which the ledger is used for integrity commitments rather than full data storage, and the platform is engineered to remain deployable under privacy, cost, and capability constraints. In practice, this points toward hybrid architectures that keep detailed evidence off chain while anchoring compact commitments on chain, higher throughput networks (e.g., Layer 2 systems) for frequent event logging, and explicit socio ethical design principles that treat usability, fairness, privacy, and safety as first class requirements rather than afterthoughts. This is consistent with \textit{Electronic Markets} research that treats blockchain value as contingent on market design, interoperability, and governance rather than simple disintermediation \citep{alt2020blockchainMarkets,bendig2025multisidedPlatforms,feulner2025intermediaryRoles,mafike2026interoperability}. In this study, TAFES provides that normative scaffold, while ADR provides a method to iteratively operationalise and validate it \citep{sein2011adr,sharma2025tafes,jahanbin20213tic}.

% ==========================================================
\section{Design principles and proof of concept development}
\label{sec:design-adr}

\noindent This section explains how the platform is designed as a socio technical artefact. TAFES is used to derive implementable requirements, and ADR is used to iteratively build and evaluate a proof of concept. The goal is to maintain a clear chain of logic from the integrity and adoption challenges in Section~\ref{sec:blockchain} to concrete design choices and evaluation checks.

\subsection{Design science framing and Action Design Research}
\noindent The study follows design science research logic: it produces a purposeful artefact (a label authentication platform) and evaluates it against relevant goals and constraints \citep{hevner2004dsr,gregor2013dsr,peffers2007dsrmethod}. Because label authentication is a multi stakeholder problem with evolving requirements and significant context constraints (privacy, capacity, governance), we adopt Action Design Research (ADR) to combine iterative building with intervention and evaluation \citep{sein2011adr,jensen2019dsr}. ADR supports iterative cycles of building and refinement. We build the artefact, demonstrate it in a bounded scenario, evaluate it using technical and stakeholder facing checks, and then refine requirements and mechanisms.

\subsection{TAFES as actionable design principles}
\label{sec:tafes-principles}

\noindent TAFES (Transparency, Accountability, Fairness, Ethics, and \textbf{Safety}) is used as a normative frame for designing a trustworthy label authentication platform \citep{sharma2025tafes}. The principles are treated jointly because certification ecosystems involve real tensions: transparency versus privacy, security versus usability, and accountability versus operational burden \citep{bernards2022veil,balzarova2020conundrum,kouhizadeh2021adoption,lautenschlager2025strikingBalance}. In this paper, \textbf{Safety} intentionally goes beyond cybersecurity to include privacy protection, operational resilience, and stakeholder harm reduction. This aligns with governance oriented work that treats decentralised platforms as value laden infrastructures rather than neutral technical components \citep{beck2018governance,frizzobarker2020disruptive,bendig2025multisidedPlatforms,feulner2025intermediaryRoles}.

\subsubsection{Transparency}
\noindent Transparency requires that stakeholders can verify label claims, provenance steps, and supporting evidence using auditable records rather than opaque assertions \citep{nikolakis2018eve,santos2021thirdparty,zavolokina2020buyersLemons,bauer2020trustedCarData}. In the proof of concept, transparency is implemented through verifiable lifecycle event logs and cryptographic links to evidence, presented in role appropriate views (consumer scan, auditor view, producer portal).

\subsubsection{Accountability}
\noindent Accountability requires that actions are attributable to authorised actors so that audits, dispute resolution, and enforcement are feasible \citep{katsikouli2020benefits,kshetri2021developing}. Where automation is used (e.g., workflow gates), accountability also requires exception handling and human override paths that preserve audit trails \citep{balzarova2020conundrum,mendling2018bpm,feulner2025intermediaryRoles}.

\subsubsection{Fairness}
\noindent Fairness requires that platform participation is not exclusionary for stakeholders with limited connectivity, limited digital skills, or limited resources \citep{kshetri2021developing,kouhizadeh2021adoption,fani2025cultivatingTrust}. This implies low friction onboarding, interfaces that work under constrained conditions, and verification access that does not require privileged intermediaries.

\subsubsection{Ethics}
\noindent Ethics requires that the platform advances fair trade outcomes in substance rather than improving documentation alone. Ethical design includes safeguards against perverse incentives, careful handling of sensitive information, and attention to whether transparency improves outcomes or merely reporting \citep{bager2022not,bernards2022veil,friedman2022sustainability,rogalski2024worthIt}.

\subsubsection{Safety}
\noindent Safety includes cybersecurity (access control, integrity checks), privacy protection, operational resilience, and minimisation of stakeholder harm. In practice this motivates hybrid on chain and off chain designs and selective disclosure so that verification does not require publishing sensitive producer or commercial data on chain \citep{kshetri2021developing,bernards2022veil,rejeb2020food,lautenschlager2025strikingBalance,mafike2026interoperability}.

\begin{figure}[ht]
 \centering
 \makebox[\textwidth][c]{\includegraphics[width=0.98\textwidth]{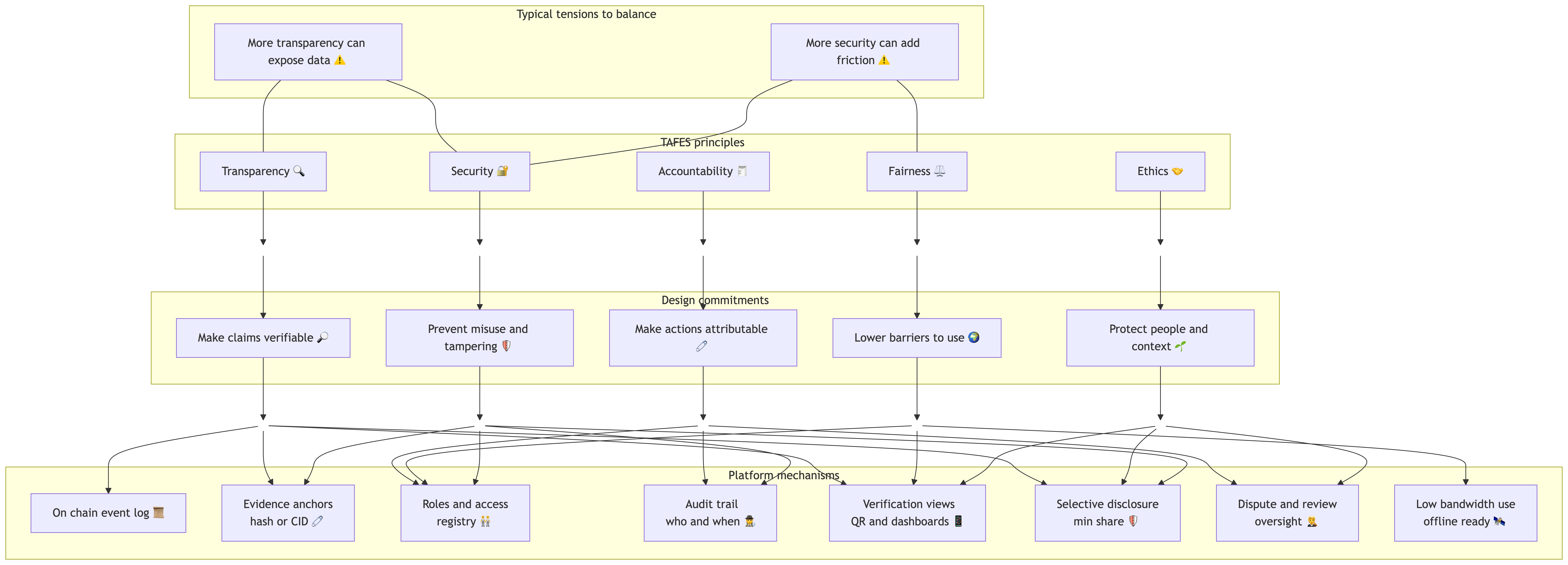}}%
 \caption{TAFES principles mapped to platform mechanisms and trade offs (e.g., transparency versus privacy; safety versus usability).}
 \label{fig:tafes_fig}
\end{figure}

\subsection{From principles to platform requirements and evaluation checks}
\label{sec:principles-to-requirements}

\noindent Table~\ref{tab:tafes-req} translates each principle into implementable requirements and example evaluation checks. This mapping provides a traceable design rationale. It explains what is built, why it is built, and what counts as evidence that the proof of concept meets its trust and adoption goals.

\begin{table}[ht]
\centering
\caption{TAFES principles translated into platform requirements and evaluation checks}
\label{tab:tafes-req}
\setlength{\tabcolsep}{2pt}\scriptsize\renewcommand{\arraystretch}{0.98}
\begin{tabular}{@{}p{0.18\textwidth}p{0.45\textwidth}p{0.29\textwidth}@{}}
\toprule
\textbf{Principle} & \textbf{Operational requirements} & \textbf{Example evaluation checks} \\
\midrule
Transparency &
Verifiable provenance chain; evidence linkage (hash/CID); readable verification views for each role &
Provenance completeness (\% batches with all lifecycle steps); evidence verifiability (CID/hash match rate); consumer task success in QR verification \\
\addlinespace
Accountability &
Role based permissions; attributable actions; tamper evident audit trails; exception and dispute mechanisms &
Unauthorised action rejection rate; audit trail completeness; dispute reconstruction feasibility from logs \\
\addlinespace
Fairness &
Low friction onboarding; constrained environment usability (low bandwidth/offline); non exclusionary verification access &
Onboarding time and failure rate by stakeholder type; usability under low bandwidth; access parity (which roles can verify what) \\
\addlinespace
Ethics &
Minimise harmful incentive shifts; protect sensitive data; align reporting with fair trade outcomes &
Stakeholder perceived legitimacy; evidence that premium distribution and claims are auditable; absence of design induced exclusion \\
\addlinespace
Safety &
Security controls (authentication and authorisation, key handling); privacy by design via hybrid data; operational resilience &
Access control tests; privacy leakage checks (sensitive data not on chain); recovery from key loss or outages \\
\bottomrule
\end{tabular}
\end{table}

\subsection{Proof of concept instantiation}
\label{sec:poc-instantiation}

\subsubsection{Hybrid Layer 2 plus off chain evidence architecture}
\noindent To address scalability, privacy, and operational feasibility, the proof of concept uses a hybrid architecture. Compact lifecycle events and integrity anchors are recorded on a scalable blockchain layer (Ethereum Layer 2), while evidence (certificates, audit reports, supporting documents) is stored off chain and referenced using content identifiers (e.g., IPFS CIDs) \citep{rejeb2020food,kostamis2021ethereumdatastores,alt2020blockchainMarkets}. This reduces on chain storage overhead while preserving verifiability through cryptographic commitment.

\subsubsection{Stakeholder roles and permissions}
\noindent The proof of concept models producers, processors, logistics actors, retailers, certifiers and validators, regulators and auditors, and consumers. Role based permissions support accountability, while read access is shaped by transparency and privacy requirements \citep{katsikouli2020benefits,kshetri2021developing,bernards2022veil,lautenschlager2025strikingBalance}. Consumer access is designed around simple QR verification to improve communicability \citep{sodamin2022fair,dionysis2022tpbcoffee,ma2025consumerQuality}.

\subsubsection{Event model and lifecycle recording}
\noindent Lifecycle steps (e.g., produced, processed, shipped, received, at retail, sold) are represented as attributable events with stable identifiers. Each event may include a pointer to off chain evidence via an integrity link. This supports audit reconstruction and dispute resolution without requiring publication of sensitive payloads on chain \citep{nikolakis2018eve,santos2021thirdparty,lautenschlager2025strikingBalance}.

\subsection{Summary}
\label{sec:design-adr-summary}

\noindent Sections~\ref{sec:blockchain}--\ref{sec:design-adr} established the integrity and adoption problem in ethical label ecosystems, reviewed blockchain affordances and constraints, and introduced a TAFES and ADR driven approach to building a proof of concept label authentication platform. The remainder of the paper provides implementation details, evaluation results, discussion of limitations and implications, and conclusions.

% ==========================================================
\section{Platform architecture and prototype implementation}
\label{sec:architecture}

\noindent Section~\ref{sec:design-adr} translated the TAFES principles into implementable requirements and evaluation checks for a label authentication platform. This section operationalises that design logic by describing the resulting system architecture and the proof of concept implementation that instantiates TAFES in a realistic, multi stakeholder setting. The emphasis here is on how the artefact is structured (layers, data model, permissions, workflow logic, and implementation components). Section~\ref{sec:results} then evaluates the proof of concept against the technical and socio ethical checks defined earlier, consistent with ADR's build--intervene--evaluate cycle \citep{sein2011adr,jensen2019dsr}.

\subsection{ADR build cycle and artefact boundary}
\noindent Following ADR, the proof of concept is developed as a socio technical artefact shaped by iterative refinement between design requirements (TAFES) and feasibility constraints (cost, latency, privacy, adoption friction) that are repeatedly highlighted in the agri food blockchain literature \citep{hilten2020blockchain,katsikouli2020benefits,kshetri2021developing,kouhizadeh2021adoption,fani2025cultivatingTrust,ma2025consumerQuality,lautenschlager2025strikingBalance}. The build phase focuses on three boundary decisions that directly connect Section~\ref{sec:blockchain} to the evaluation in Section~\ref{sec:results}.

\begin{enumerate}
    \item \textbf{Evidence centred provenance:} the platform treats provenance as a chain of lifecycle events, each of which can be linked to supporting evidence (certificates, reports, audits). Sensitive payloads remain off chain to address privacy and regulatory constraints \citep{bernards2022veil,balzarova2020conundrum,alt2020blockchainMarkets}.
    \item \textbf{Low friction user experience with mediated blockchain interaction:} to mitigate consumer comprehension barriers and reduce onboarding friction, blockchain interactions are mediated by the platform backend and presented through familiar interfaces (web or mobile, QR scanning). This reflects findings that usability and communicability are prerequisites for trust gains \citep{sodamin2022fair,dionysis2022tpbcoffee,contini2023credence,ma2025consumerQuality}.
    \item \textbf{Scalable anchoring through a hybrid design:} the proof of concept anchors compact integrity commitments on chain and stores detailed evidence off chain. This follows common design guidance for reducing on chain storage burden while preserving verifiability \citep{rejeb2020food,sanka2021scalability,kostamis2021ethereumdatastores,alt2020blockchainMarkets}.
\end{enumerate}

\noindent The artefact is demonstrated using a coffee supply chain scenario because coffee is a flagship fair trade category with strong consumer awareness and frequent discussion in blockchain for fairness research \citep{bager2022not,owsianowski2025linking,samoggia2025promised}. The scenario supports end to end evaluation by checking whether provenance is complete, whether evidence can be independently verified, and whether accountability can be traced across stakeholders, while remaining sufficiently bounded for repeatable benchmarking in Section~\ref{sec:results}.

\begin{figure}[ht]
 \centering
 \makebox[\textwidth][c]{\includegraphics[width=0.98\textwidth]{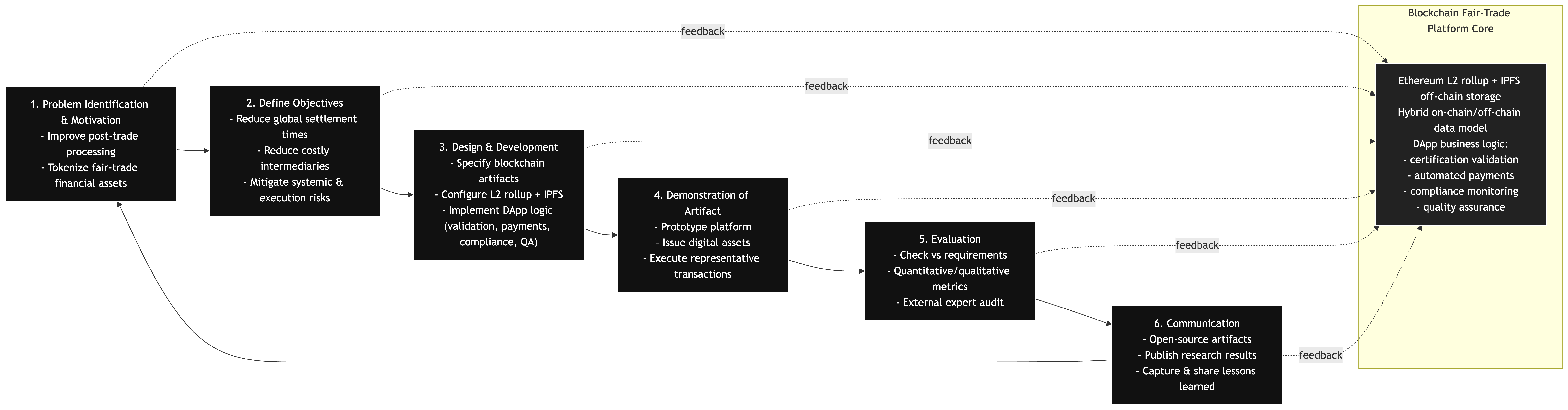}}%
 \caption{ADR build logic adopted in the proof of concept: problem framing, objectives, design and development, demonstration, evaluation, and communication, with iterative feedback loops \citep{sein2011adr,jensen2019dsr}.}
 \label{fig:adr_stages}
\end{figure}

\subsection{Layered platform architecture}
\begin{figure}[ht]
 \centering
 \makebox[\textwidth][c]{\includegraphics[width=0.98\textwidth]{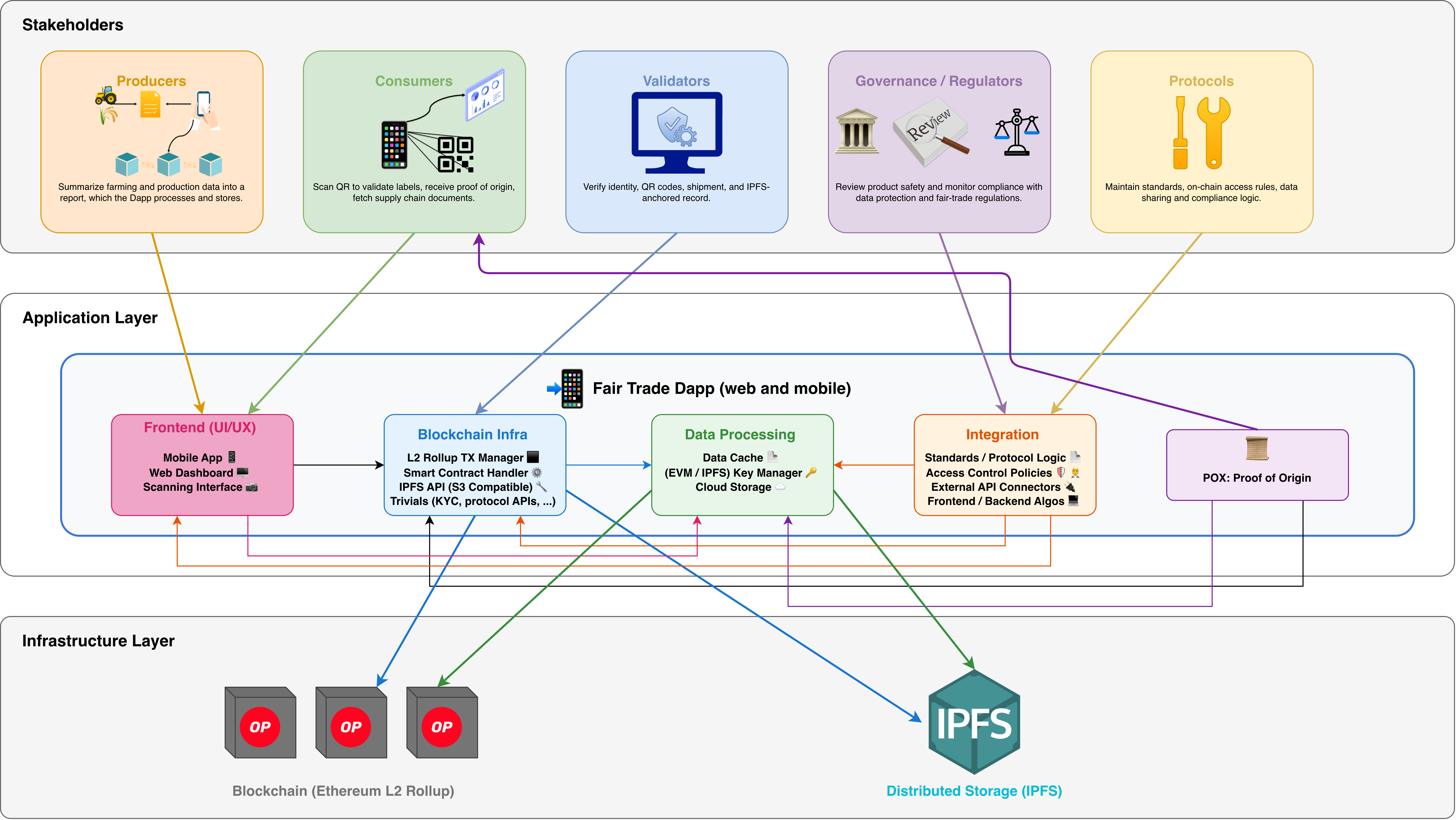}}%
 \caption{Proof of concept platform architecture. Stakeholders interact with a Fair Trade DApp (web or mobile). Evidence is stored off chain and anchored on chain through compact identifiers to provide auditable, tamper evident provenance.}
 \label{fig:arch_overview}
\end{figure}

\noindent Figure~\ref{fig:arch_overview} presents the proof of concept architecture as three interacting layers. Stakeholders form the ecosystem layer; the application layer implements TAFES aligned workflows; and the infrastructure layer provides immutable anchoring and evidence availability. This layered structure is motivated by the need to support heterogeneous stakeholder capabilities and incentives, minimise operational burden, and avoid re centralising trust in a single organisation while remaining practically deployable \citep{hilten2020blockchain,kouhizadeh2021adoption,kshetri2021developing,beck2018governance,bendig2025multisidedPlatforms,feulner2025intermediaryRoles}.

\subsubsection{Stakeholder layer}
\noindent The proof of concept models the stakeholder categories commonly emphasised in fair trade certification ecosystems: producers, processors, retailers and distributors, certifiers and validators, regulators and auditors, and consumers. The design treats these actors as co producers of provenance. Each contributes lifecycle events and or verification actions, enabling accountability and dispute reconstruction without relying exclusively on periodic audits \citep{katsikouli2020benefits,santos2021thirdparty,nikolakis2018eve,fani2025cultivatingTrust}. Consumer interaction is intentionally lightweight (QR based verification) because consumer studies consistently show that trust gains depend on low effort access to understandable evidence \citep{sodamin2022fair,dionysis2022tpbcoffee,lou2024denim,ma2025consumerQuality}.

\subsubsection{Application layer (Fair Trade DApp and service components)}
\noindent The application layer implements the business logic required to operationalise TAFES:
\begin{itemize}
    \item \textbf{Transparency mechanisms:} user facing provenance views and evidence retrieval pathways that clarify what can be verified and what remains private.
    \item \textbf{Accountability mechanisms:} role based action permissions, attributable event submission, and auditable logs that support certification checks and disputes.
    \item \textbf{Fairness mechanisms:} simplified workflows that do not require stakeholders to run blockchain nodes or manage complex key material, reducing exclusion risk in constrained environments \citep{kshetri2021developing,kouhizadeh2021adoption,fani2025cultivatingTrust}.
    \item \textbf{Ethics and safety mechanisms:} privacy by design choices (minimal on chain data, selective disclosure) and operational safeguards to reduce stakeholder harm from over exposure of sensitive commercial or personal information \citep{bernards2022veil,balzarova2020conundrum,friedman2022sustainability,lautenschlager2025strikingBalance}.
\end{itemize}

\subsubsection{Infrastructure layer (hybrid anchoring and evidence storage)}
\noindent The infrastructure layer provides two complementary capabilities. It uses a blockchain substrate for tamper evident lifecycle anchoring, and it uses an off chain evidence store for documents and media linked to those events. This responds directly to the scalability and privacy constraints of public ledgers in high frequency provenance systems \citep{rejeb2020food,sanka2021scalability,kostamis2021ethereumdatastores,alt2020blockchainMarkets,mafike2026interoperability}. Evidence objects are referenced using content identifiers (CIDs) in a content addressed store (e.g., IPFS), enabling integrity verification without placing the evidence payload itself on chain \citep{benet2014ipfs}.

\subsection{Evidence centric data model and CID anchoring}
\noindent The proof of concept uses an evidence centric model in which each supply chain event can be associated with one or more evidence objects (e.g., certification documents, inspection reports, shipment records). The platform anchors a compact commitment on chain that binds together a batch identifier, a lifecycle step identifier, and an evidence CID (or hash) that can be independently verified.

\noindent This design makes two TAFES driven trade offs explicit. First, transparency is delivered through verifiable commitments, while confidentiality is protected by keeping evidence payloads off chain \citep{bernards2022veil,balzarova2020conundrum,lautenschlager2025strikingBalance}. Second, auditability is preserved through compact anchoring, while cost is controlled by avoiding on chain storage growth with document size \citep{rejeb2020food,sanka2021scalability,alt2020blockchainMarkets}.

\begin{figure}[ht]
 \centering
 \makebox[\textwidth][c]{\includegraphics[width=0.92\textwidth]{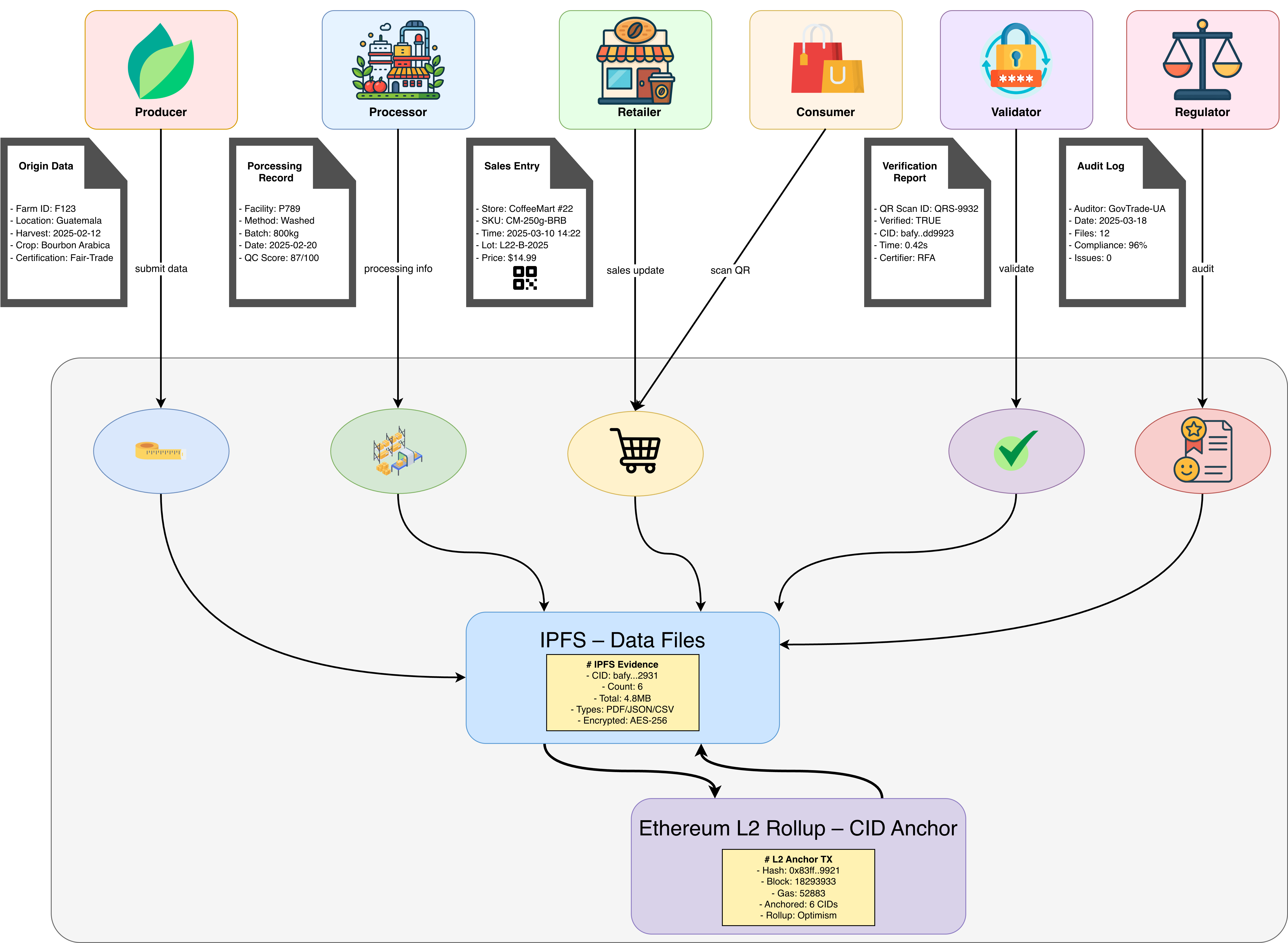}}%
 \caption{Evidence and anchoring model. Stakeholders submit events and evidence; evidence is stored off chain as data files and referenced by content identifiers. Compact CID anchors are recorded on an Ethereum Layer 2 to provide tamper evident verification.}
 \label{fig:cid_flow}
\end{figure}

\noindent In operational terms, the proof of concept separates three categories of information. Event metadata supports audit reconstruction (who acted, what occurred, when it occurred, and which lifecycle step was affected). Evidence pointers (CIDs or hashes) provide integrity commitments that can be verified independently. Evidence payloads (documents and media) remain off chain with controlled access paths. Only event metadata and evidence pointers are anchored on chain, while evidence payloads remain off chain. This reflects widely noted design implications for deploying blockchain in certification ecosystems: the ledger should protect integrity, while governance and data access policies manage confidentiality \citep{cao2021governance,kshetri2021developing,alt2020blockchainMarkets,feulner2025intermediaryRoles}.

\subsection{Stakeholder permissions and accountability paths}
\noindent To operationalise accountability (Table~\ref{tab:tafes-req}), the proof of concept models stakeholder actions as role authorised operations. The platform uses role based permissions so that write access is constrained to actors responsible for each lifecycle step, while read access is shaped by verification needs and privacy constraints \citep{katsikouli2020benefits,bernards2022veil,lautenschlager2025strikingBalance}. Table~\ref{tab:roles_actions} summarises the role model used in implementation and evaluation.

\begin{table}[ht]
\centering
\caption{Proof of concept stakeholder roles, primary actions, and provenance responsibilities}
\label{tab:roles_actions}
\setlength{\tabcolsep}{2pt}\scriptsize\renewcommand{\arraystretch}{0.98}
\begin{tabular}{@{}p{0.22\textwidth}p{0.42\textwidth}p{0.28\textwidth}@{}}
\toprule
\textbf{Role} & \textbf{Primary platform actions} & \textbf{Accountability contribution} \\
\midrule
Producer &
Create batch; submit origin and production event; attach evidence (e.g., cooperative records, certificates) &
Establishes the provenance root; accountable for initial claims (oracle risk managed via verification and gating) \\
\addlinespace
Processor &
Submit processing events; attach processing and quality evidence &
Links transformations to the source batch; supports traceability across value added stages \\
\addlinespace
Retailer/Distributor &
Submit receipt and retail readiness events; record sale; generate and print QR for consumer verification &
Bridges the operational chain to the point of sale; supports consumer facing verification \\
\addlinespace
Certifier / Validator &
Verify or attest compliance; attach inspection or audit evidence; approve or reject gating conditions &
Adds independent verification and dispute relevant attestations \citep{santos2021thirdparty,feulner2025intermediaryRoles} \\
\addlinespace
Regulator / Auditor &
Read only oversight; periodic review; request or verify evidence for compliance purposes &
Strengthens governance; supports external accountability pathways \citep{bernards2022veil} \\
\addlinespace
Consumer &
Scan QR; view verification summary; optionally provide feedback or dispute signal &
Operationalises transparency at purchase time; converts provenance into communicable trust \citep{sodamin2022fair,contini2023credence,ma2025consumerQuality} \\
\bottomrule
\end{tabular}
\end{table}

\noindent The model explicitly recognises the oracle problem. Producers and intermediaries can submit inaccurate information even if it becomes immutable after anchoring \citep{caldarelli2020oracle,hassan2023oracle}. The proof of concept therefore structures verification as a combination of role accountability, validator attestations, and evidence linkage that enables after the fact audits and dispute resolution. This operationalises TAFES by treating data integrity as both a technical and governance problem \citep{balzarova2020conundrum,friedman2022sustainability,beck2018governance,feulner2025intermediaryRoles}.

\subsection{Supply chain workflow and transaction design}
\noindent The proof of concept implements a lifecycle model aligned with common fair trade traceability narratives in the coffee sector \citep{bager2022not,owsianowski2025linking,samoggia2025promised}. A traceable batch progresses through six coarse grained steps:
\texttt{Produced}, \texttt{Processed}, \texttt{Shipped}, \texttt{Received}, \texttt{AtRetail}, and \texttt{Sold}.
At each step, an authorised actor submits an event and, where appropriate, anchors evidence pointers. This produces an auditable chain of custody that can be reconstructed by validators and presented to consumers through a QR scan, addressing the credibility and trust issues highlighted in Section~\ref{sec:blockchain}.

\begin{figure}[ht]
 \centering
 \makebox[\textwidth][c]{\includegraphics[width=0.6\linewidth,keepaspectratio]{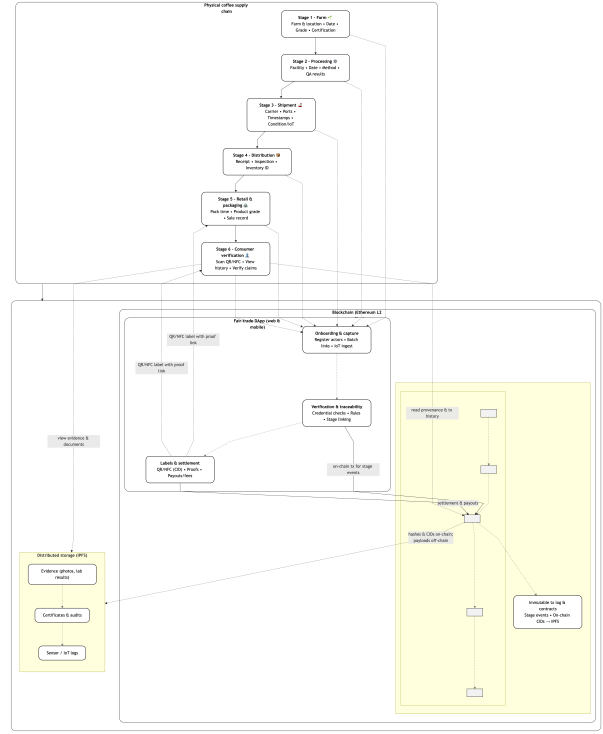}}%
 \caption{Process flow used in the proof of concept coffee scenario: provenance events and evidence are generated across farm production, processing, shipment, distribution and retail, and consumer verification stages.}
 \label{fig:coffee_process}
\end{figure}

\begin{figure}[ht]
 \centering
 \makebox[\textwidth][c]{\includegraphics[width=0.90\textwidth]{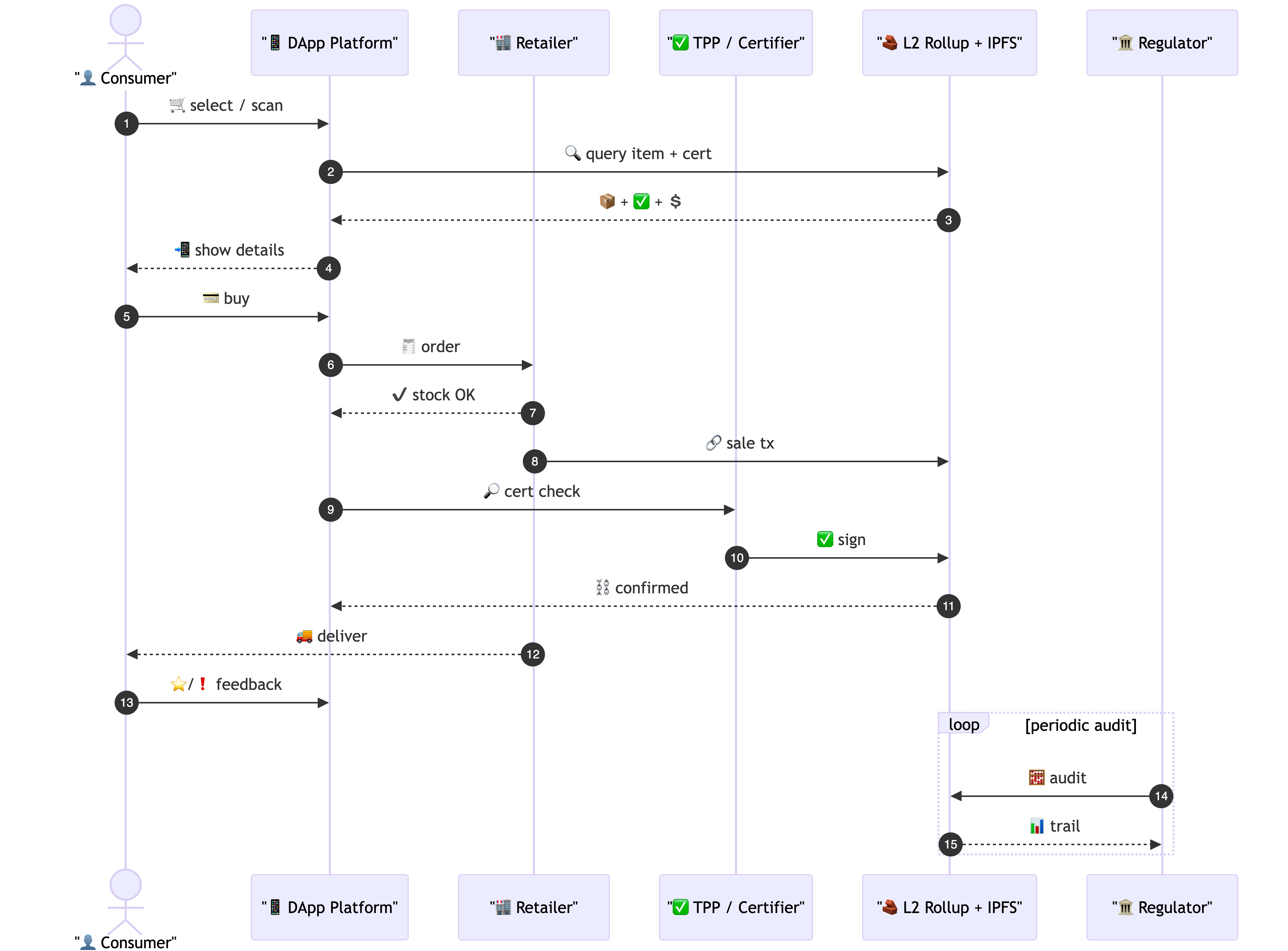}}%
 \caption{Consumer facing interaction sequence in the proof of concept: selection and scan, verification retrieval, purchase recording, certification check and attestation, and optional regulatory audit trail.}
 \label{fig:sequence_diagram}
\end{figure}

\noindent In addition to the stage model (Figure~\ref{fig:coffee_process}), the proof of concept includes a consumer facing interaction sequence (Figure~\ref{fig:sequence_diagram}) to operationalise communicable verification \citep{sodamin2022fair,dionysis2022tpbcoffee,ma2025consumerQuality}. The design goal is not to teach blockchain concepts to consumers, but to provide a verification outcome that is easily interpretable: whether a product's label claim is supported by a complete, attested provenance chain and whether supporting evidence can be fetched and validated.

\subsection{Prototype implementation components}
\noindent The proof of concept is implemented as a modular stack with separation between on chain integrity and workflow enforcement, off chain evidence storage and retrieval, and user facing applications. This separation is consistent with prior work emphasising that successful supply chain blockchain systems require careful integration of on chain and off chain governance and service components \citep{cao2021governance,rejeb2020food,alt2020blockchainMarkets,mafike2026interoperability}.

\subsubsection{On chain modules (smart contracts)}
\noindent The smart contract layer encodes safety critical logic: role checks, lifecycle state transitions, and event anchoring. Contracts are organised by responsibility (role management, evidence anchoring, event batching, lifecycle management, and optional payment routing). This mirrors patterns proposed for certification automation using smart contracts and process aware blockchain systems \citep{santos2021thirdparty,nikolakis2018eve,mendling2018bpm,lautenschlager2025strikingBalance}. Contract names, interfaces, and on chain checks used in evaluation are documented in Appendix~\ref{app:smart_contracts}.

\subsubsection{Off chain services and blockchain as a service operation}
\noindent To reduce operational burden for ecosystem participants, the proof of concept is deployed in a service oriented manner. Stakeholders interact through web or mobile interfaces, while the backend manages blockchain interactions (transaction submission, batching, and indexing) and evidence operations (uploading, pinning, and retrieval). This approach directly addresses adoption barriers in multi stakeholder settings where participants may have limited technical capability or infrastructure \citep{kshetri2021developing,kouhizadeh2021adoption,hilten2020blockchain,nayal2023antecedents,fani2025cultivatingTrust,feulner2025intermediaryRoles}. It also supports TAFES fairness by avoiding exclusion based on the ability to operate nodes or manage complex blockchain tooling.

\subsubsection{User interfaces}
\noindent The proof of concept provides role appropriate interfaces. Producers and processors use data entry views designed for constrained environments. Retailers use views that support QR generation and inventory linked provenance. Validators and regulators use views that support evidence inspection and audit reconstruction. Consumers use a QR scan experience that summarises verification status and provides optional drill down into provenance and evidence. This aligns with evidence that consumer facing verification must be accessible and simple to translate provenance into trust and purchase intention \citep{contini2023credence,dionysis2022tpbcoffee,lou2024denim,ma2025consumerQuality}.

\subsection{Instrumentation and evaluation readiness}
\noindent Section~\ref{sec:results} evaluates the proof of concept using a mixed set of checks derived from Table~\ref{tab:tafes-req}. To ensure the evaluation is reproducible and directly attributable to design choices, the proof of concept includes instrumentation that logs:

\begin{itemize}
    \item \textbf{Technical performance:} event submission latency, anchoring throughput under load, and resource and cost proxies (e.g., per event on chain operations) \citep{sanka2021scalability,kostamis2021ethereumdatastores}.
    \item \textbf{Provenance integrity:} completeness of lifecycle steps per batch and verifiability of evidence pointers (CID or hash match and retrievability) \citep{nikolakis2018eve,santos2021thirdparty}.
    \item \textbf{Accountability enforcement:} rejection of unauthorised actions, audit reconstruction feasibility, and traceability of actor attributed submissions \citep{katsikouli2020benefits,bernards2022veil}.
    \item \textbf{Communicability (consumer facing):} task based verification success in QR based flows, reflecting the adoption emphasis in consumer studies \citep{sodamin2022fair,contini2023credence,ma2025consumerQuality}.
\end{itemize}

% ==========================================================
\section{Evaluation and results}
\label{sec:results}

\noindent This section evaluates the proof of concept against the TAFES derived requirements in Table~\ref{tab:tafes-req} and the research questions in Section~\ref{sec:introduction}. Following ADR logic, the evaluation combines empirical measurements on the selected blockchain and evidence substrates with model based stress analyses parameterised by the measured results. The intent is to provide a clear chain of reasoning from the design choices in Section~\ref{sec:architecture} to observable behaviour (Transparency and Accountability), operational feasibility (Fairness and Safety), and the boundary conditions under which those claims remain valid.

\subsection{Evaluation setting and experiment configuration}
\label{sec:eval_env}

\noindent The proof of concept targets a hybrid architecture in which integrity anchors are recorded on an Ethereum Layer 2 network and supporting evidence is stored off chain in a content addressed store. The evaluation therefore separates on chain inclusion behaviour from off chain evidence availability. This separation is necessary because end to end verification is constrained by the slower path in a given user flow. For example, consumer QR verification depends on prompt query responses and evidence retrieval, while auditability depends on reconstructable logs and verifiable evidence integrity.

\begin{table}[ht]
\centering
\caption{Evaluation setting and configuration used in reference runs}
\label{tab:eval_env}
\setlength{\tabcolsep}{2pt}\scriptsize\renewcommand{\arraystretch}{0.98}
\begin{tabular}{@{}p{0.28\textwidth}p{0.66\textwidth}@{}}
\toprule
\textbf{Dimension} & \textbf{Setting / definition used} \\
\midrule
Blockchain network &
Optimism Sepolia (OP Sepolia), Chain ID \texttt{11155420}. Confirmation is defined as transaction inclusion in an L2 block (inclusion confirmation, not economic finality). OP Stack chains target a short block interval (nominally $\approx 2$~s), but observed inclusion latency varies with network load and RPC behaviour. \citep{optimismSepoliaDocs,opStackGlossary} \\
\addlinespace
On chain implementation &
The proof of concept uses a modular smart contract suite that implements role management, lifecycle transitions, and batched anchoring of evidence commitments. Contract names and interface details are provided in Appendix~\ref{app:smart_contracts}. \\
\addlinespace
RPC providers &
Public and private OP Sepolia RPC endpoints supplied via environment configuration. Benchmark scripts support multi provider comparisons and multi wallet submission to probe rate limiting and effective throughput. \\
\addlinespace
Tooling and harness &
The evaluation harness is implemented in a TypeScript and Node.js environment and uses a standard Ethereum development framework for contract interaction. Implementation specific package and version details are documented in Appendix~\ref{app:reproducibility}. \\
\addlinespace
Evidence store (off chain) &
An IPFS compatible RPC endpoint (Filebase) is used as the evidence substrate with pinning enabled. Evidence objects are added and retrieved through standard IPFS RPC operations; endpoint level details are provided in Appendix~\ref{app:reproducibility}. \\
\addlinespace
Trial counts ($n$) &
Evidence verifiability experiment: four file sizes (10~KB, 100~KB, 1~MB, 5~MB) with 10 repeats each, yielding $n=40$ evidence objects. Each object was uploaded, fetched, and CID verified. \\
\addlinespace
Execution host &
A local workstation running macOS executed the evaluation harness and benchmark runs. \\
\bottomrule
\end{tabular}
\end{table}

\noindent The reference runs above establish the empirical basis for Sections~\ref{sec:results_rq3}--\ref{sec:results_accountability}. Sections~\ref{sec:stress_batching}--\ref{sec:stress_oracle_privacy} then extend the empirical findings through stress analyses grounded in measured throughput, latency, and gas profiles.

\subsection{Evaluation design and evidence sources (ADR alignment)}
\label{sec:eval_strategy}

\noindent The evaluation is organised into complementary strands that map directly to the TAFES operational checks (Table~\ref{tab:tafes-req}) and the adoption barriers identified in Section~\ref{sec:blockchain}. Table~\ref{tab:eval_overview} summarises the strands and the evidence they produce.

\begin{table}[ht]
\centering
\caption{Evaluation strands, evidence sources, and mapping to TAFES requirements}
\label{tab:eval_overview}
\setlength{\tabcolsep}{2pt}\scriptsize\renewcommand{\arraystretch}{0.98}
\begin{tabular}{@{}p{0.21\textwidth}p{0.44\textwidth}p{0.28\textwidth}@{}}
\toprule
\textbf{Evaluation strand} & \textbf{Evidence source / procedure} & \textbf{Primary TAFES checks} \\
\midrule
On chain anchoring benchmark &
OP Sepolia load tests of batched anchoring for evidence commitments; measure inclusion latency, peak event rate, maximum batch size, and gas usage per commitment &
Safety (cost viability), Transparency (auditability under load) \\
\addlinespace
End to end scenario execution &
Coffee batch lifecycle walk through (\texttt{Produced} $\rightarrow$ \texttt{Sold}) with evidence commitments anchored at each step &
Transparency (completeness, verifiability), Accountability (traceability) \\
\addlinespace
Accountability enforcement &
Negative tests (unauthorised writes, replay attempts), plus audit reconstruction from logs and linked evidence; query latency simulation for reconstruction &
Accountability (authorised writes), Safety (tamper evidence, replay resistance) \\
\addlinespace
Evidence verifiability &
Upload, fetch, and CID verification experiment against the IPFS RPC configuration, capturing success rates and latency distributions &
Transparency and Safety (retrievability, CID integrity) \\
\addlinespace
Model based stress analyses &
Batching delay model, fee sensitivity rescaling, evidence availability under churn, and oracle and audit sampling sensitivity parameterised by measured results &
Safety and Fairness (operational feasibility), Accountability (detectability), Transparency (user perceived verification latency) \\
\bottomrule
\end{tabular}
\end{table}

\subsection{Metrics and operational definitions}
\label{sec:metrics}

\noindent The core integrity metrics operationalise the requirements in Table~\ref{tab:tafes-req}. Provenance completeness captures the fraction of evaluated batches that contain all required lifecycle steps. Evidence retrievability captures the fraction of evidence objects that can be fetched by CID. CID match captures whether retrieved bytes recompute to the expected CID, supporting integrity verification. Evidence verifiability captures the overall fraction of evaluated evidence objects that are both retrievable and integrity consistent.

\noindent To define these metrics compactly, let $N_B$ denote the number of evaluated batches and $N_B^{\mathrm{full}}$ denote the number of those batches that contain all required lifecycle steps. Let $N_E$ denote the number of evaluated evidence objects, $N_E^{\mathrm{fetch}}$ the number fetched successfully by CID, and $N_E^{\mathrm{cid}}$ the number whose retrieved bytes recompute to the expected CID. We then define:

\begin{equation}
\label{eq:core_metrics}
\begin{aligned}
C \; &= \; \frac{N_B^{\mathrm{full}}}{N_B} \qquad\qquad\qquad\;\;\;\;\;\;\;\; \text{(provenance completeness)} \\
R \; &= \; \frac{N_E^{\mathrm{fetch}}}{N_E} \qquad\qquad\qquad\qquad\qquad \text{(retrievability rate)} \\
M \; &= \; \frac{N_E^{\mathrm{cid}}}{N_E^{\mathrm{fetch}}} \qquad\qquad\qquad\qquad\;\; \text{(CID match rate)} \\
V \; &= \; \frac{N_E^{\mathrm{cid}}}{N_E} \;=\; R \cdot M \qquad\qquad\qquad\;\; \text{(evidence verifiability)}
\end{aligned}
\end{equation}

\noindent Because user perceived verification depends on timeliness, we define an anchoring delay for application events:

\begin{equation}
\label{eq:e2e_delay}
D_{\text{anchor}} = t_{\text{included}} - t_{\text{arrive}}
\end{equation}

\noindent For auditability, we retain the end to end Audit Query Latency (AQL) metric used in the proof of concept evaluation harness:

\begin{equation}
\label{eq:aql}
\mathrm{AQL} = T_{\text{receipts}} + T_{\text{decode}} + T_{\text{sort}} + T_{\text{timestamps}}
\end{equation}

\noindent where $T_{\text{receipts}}$ is the time to fetch transaction receipts, $T_{\text{timestamps}}$ is the time to attach block timestamps, and the remaining terms are local decode and ordering costs.

\subsection{Results: on chain anchoring throughput and cost profile (RQ3)}
\label{sec:results_rq3}

\noindent RQ3 concerns whether frequent anchoring of provenance and evidence commitments can be achieved with sufficiently low latency and marginal cost to remain viable for low margin supply chains. In the proof of concept, detailed evidence remains off chain and the chain is used primarily for compact commitments (CIDs and minimal event metadata) rather than storage heavy updates (Figure~\ref{fig:cid_flow}). This design choice follows prior arguments that blockchain market infrastructures often combine on-chain and off-chain data components to balance transparency, privacy, and operational cost \citep{alt2020blockchainMarkets,mafike2026interoperability}.

\noindent \textit{Benchmark procedure.} Reference measurements were obtained on OP Sepolia using a benchmark harness that submits high density batched anchoring transactions across multiple funded accounts and RPC providers. The harness logs per transaction gas usage, inclusion delay, and the largest batch size that confirms without out of gas reverts (identified using a binary search strategy). Throughput is reported as application level commitments per second because the operational workload is provenance and evidence anchoring rather than raw transaction count. Implementation details for the benchmarking harness are provided in Appendix~\ref{app:reproducibility}.

\begin{table}[ht]
\centering
\caption{On chain benchmarking results for CID based anchoring on OP Sepolia (empirical)}
\label{tab:op_bench}
\setlength{\tabcolsep}{2pt}\scriptsize\renewcommand{\arraystretch}{0.98}
\begin{tabular}{@{}p{0.33\textwidth}p{0.18\textwidth}p{0.42\textwidth}@{}}
\toprule
\textbf{Metric} & \textbf{Observed value} & \textbf{Interpretation for TAFES / RQ3} \\
\midrule
Peak realised anchoring throughput &
$\approx 1{,}400$ CID commitments/s &
Provides headroom for dense provenance logging; reduces the risk that Transparency becomes a performance bottleneck \\
\addlinespace
Typical inclusion latency (L2 block inclusion) &
$\approx 2$ s &
Consistent with near real time anchoring needs for operational traceability and prompt auditability \\
\addlinespace
Maximum confirmed batch size (single transaction) &
$1{,}022$ CID commitments &
Upper bound for event density per transaction; batching improves throughput per unit overhead \\
\addlinespace
Gas per CID commitment (mean) &
$62{,}841$ gas &
Stable marginal gas profile supports predictable operating expenditure \\
\addlinespace
Reference lifecycle definition &
$6$ coarse steps, $13$ CID commitments &
Represents an end to end coffee batch trace with multiple evidence commitments across lifecycle steps \\
\bottomrule
\end{tabular}
\end{table}

\noindent \textit{Cost model.} For transparency, we report cost as a function of observed gas usage and an assumed effective gas price $g$ (in gwei). The estimated per batch anchoring cost is:

\begin{equation}
\label{eq:cost_batch}
C_{\text{batch}}(\mathrm{USD}) = G_{\text{batch}} \cdot g \cdot 10^{-9}\ \mathrm{ETH/gas}\ \cdot P_{\mathrm{ETH}\rightarrow \mathrm{USD}}
\end{equation}

\noindent where $G_{\text{batch}}$ is the observed gas used for anchoring one batch worth of CID commitments. In the reference run, $G_{\text{batch}}\approx 816{,}939$ gas (13 commitments $\times$ 62{,}841 gas/commitment). Because fee regimes vary materially across networks and time, we present a sensitivity table rather than a single point estimate.

\begin{table}[ht]
\centering
\caption{Scenario analysis: per batch anchoring cost under varying effective gas prices (deterministic rescaling of observed gas usage)}
\label{tab:cost_sensitivity}
\setlength{\tabcolsep}{2pt}\scriptsize\renewcommand{\arraystretch}{0.98}
\begin{tabular}{@{}p{0.19\textwidth}p{0.22\textwidth}p{0.22\textwidth}p{0.29\textwidth}@{}}
\toprule
\textbf{Gas price (gwei)} & \textbf{Cost per batch (USD)} & \textbf{Cost per CID (USD)} & \textbf{Interpretation} \\
\midrule
0.001 & 0.00151 & 0.000116 & Very low fee regimes or test conditions \\
0.01  & 0.0151  & 0.00117  & Low fee Layer 2 conditions \\
0.1   & 0.151   & 0.0117   & Moderate fee regimes; still sub dollar for batch level anchoring \\
0.5   & 0.757   & 0.0583   & Higher fee regimes; batching remains important \\
1.0   & 1.51    & 0.116    & Conservative stress case for sensitivity analysis \\
\bottomrule
\end{tabular}
\end{table}

\noindent The sensitivity analysis supports the RQ3 claim at the batch level. Because a batch typically represents multiple retail units, the per unit traceability cost remains small even when the fee regime becomes less favourable. Moreover, the proof of concept design avoids placing evidence payloads on chain, preventing cost growth with document size.

\subsection{Results: evidence retrievability and CID integrity (RQ1/RQ3)}
\label{sec:results_evidence}

\noindent Hybrid architectures are only trustworthy if off chain evidence remains retrievable and integrity verifiable by independent parties. We therefore evaluate the full evidence loop: generate bytes, compute the expected CID, upload the object, fetch by CID, and recompute the CID from retrieved bytes.

\noindent \textit{Experiment.} The evaluation harness executes $n=40$ upload, fetch, and verify trials (Table~\ref{tab:eval_env}) against the IPFS compatible RPC endpoint with pinning enabled. It records per trial latencies and produces a machine readable report. Reproducibility details, including the exact harness entry points and repository paths, are provided in Appendix~\ref{app:reproducibility}.

\begin{table}[ht]
\centering
\caption{Evidence verifiability results via IPFS RPC (empirical reference run)}
\label{tab:evidence_results}
\setlength{\tabcolsep}{2pt}\scriptsize\renewcommand{\arraystretch}{0.98}
\begin{tabular}{@{}p{0.32\textwidth}p{0.18\textwidth}p{0.42\textwidth}@{}}
\toprule
\textbf{Metric} & \textbf{Observed value} & \textbf{Interpretation for RQ1/TAFES} \\
\midrule
Evidence objects evaluated ($n$) &
40 &
Four file sizes (10~KB, 100~KB, 1~MB, 5~MB), 10 repeats each \\
\addlinespace
Retrievability rate $R$ (Eq.~\ref{eq:core_metrics}) &
1.00 &
All CIDs were fetched successfully under the pinned configuration \\
\addlinespace
CID match rate $M$ (Eq.~\ref{eq:core_metrics}) &
1.00 &
All fetched bytes recomputed to the expected CID, supporting tamper evidence \\
\addlinespace
Upload latency (ms) &
p50 = 737; p95 = 5237 &
Small objects publish quickly; tail latency increases for larger objects \\
\addlinespace
Fetch latency (ms) &
p50 = 361; p95 = 770 &
CID based retrieval remains sub second at median under test conditions \\
\addlinespace
Failures &
0 &
No timeouts, CID mismatches, or gateway errors in the reference run \\
\bottomrule
\end{tabular}
\end{table}

\noindent With 40 out of 40 successes, the observed rates yield $V=1.0$ for this reference configuration (Eq.~\ref{eq:core_metrics}). However, because the sample is finite, the implied per object success probability is bounded below by a conservative binomial confidence bound. For example, a one sided 95\% lower bound for a success probability with 40 successes in 40 trials is approximately 0.91. This motivates explicit churn and availability stress analysis in Section~\ref{sec:stress_churn} rather than treating perfect reference run performance as a universal guarantee.

\subsection{Results: provenance completeness (RQ1)}
\label{sec:results_completeness}

\noindent In the coffee scenario (Figure~\ref{fig:coffee_process}), a batch progresses through six required lifecycle steps (\texttt{Produced} $\rightarrow$ \texttt{Sold}). For the reference end to end scenario execution used in the proof of concept demonstration, each required step was recorded and linked to at least one evidence commitment. Therefore, completeness is $C=1.0$ for the demonstrated batch (Eq.~\ref{eq:core_metrics}).

\noindent Importantly, completeness is not a purely technical property. Blockchains provide tamper evidence after data is recorded; they do not guarantee that stakeholders will record every required step (the oracle and compliance dimension) \citep{caldarelli2020oracle,hassan2023oracle}. For that reason, the completeness result should be interpreted as a capability demonstration: the workflow can capture and verify all steps. Sections~\ref{sec:stress_batching} and \ref{sec:stress_oracle_privacy} therefore examine the operational and incentive conditions under which completeness remains a stable outcome.

\subsection{Results: accountability enforcement and audit reconstruction (RQ2)}
\label{sec:results_accountability}

\noindent Accountability requires that actions are attributable to authorised roles and that an auditor can reconstruct what occurred from durable records \citep{bernards2022veil,katsikouli2020benefits,feulner2025intermediaryRoles}. The proof of concept operationalises accountability through three classes of mechanisms. First, actor registration and status control restrict write access to authorised participants. Second, anchoring operations enforce role and status checks so that inactive or unauthorised accounts cannot submit events. Third, replay protection prevents duplicate anchoring of the same lifecycle step for a given product identifier. The specific contract components and state checks are documented in Appendix~\ref{app:smart_contracts}.

\subsubsection{Negative case enforcement (authorisation and replay resistance)}
\noindent The following negative cases are expected to be rejected deterministically by the deployed on chain logic:
\begin{itemize}
    \item \textbf{Unauthorised calls (not registered):} anchoring operations require an active, registered actor. Calls from unregistered accounts fail authorisation checks.
    \item \textbf{Suspended or revoked actor calls:} accounts that are not in an active status cannot submit anchoring operations.
    \item \textbf{Replay attempts:} attempts to anchor the same product identifier and lifecycle step more than once are rejected by replay protection.
    \item \textbf{Concurrent collisions:} if two actors race to anchor the same product identifier and lifecycle step, only the first included transaction succeeds; later submissions are rejected.
\end{itemize}
\noindent These controls support Accountability and Safety without relying on off chain trust assumptions.

\subsubsection{Audit reconstruction performance (AQL)}
\noindent For auditability, the proof of concept emits indexed events from the on chain components that govern lifecycle transitions and evidence commitments. An auditor reconstructs an ordered timeline by fetching relevant transaction receipts, decoding event logs, ordering records by block number and log index, and attaching timestamps. Evidence integrity is then validated by retrieving evidence objects via CID and recomputing the CID from retrieved bytes.

\noindent The audit query latency simulation repeatedly reconstructs one batch containing 10 transactions, 15 decoded events, and 10 unique blocks. The simulation runs $n=30$ trials (after 3 warm up runs) with concurrency 8 and reports p50, p95, mean, and max. Two regimes are reported: \textbf{UNCACHED} (no caches, representative of live RPC queries) and \textbf{CACHED} (receipt and timestamp caches pre populated, representative of indexer or cache backed operation). The evaluation harness details are provided in Appendix~\ref{app:reproducibility}.

\begin{table}[ht]
\centering
\caption{Audit Query Latency (AQL) for reconstructing one batch (OP Sepolia)}
\label{tab:audit_query_latency_strong_conservative}
\setlength{\tabcolsep}{2pt}\scriptsize\renewcommand{\arraystretch}{0.98}
\begin{tabular}{@{}p{0.25\textwidth}p{0.18\textwidth}p{0.18\textwidth}p{0.31\textwidth}@{}}
\toprule
\textbf{Metric} & \textbf{UNCACHED (ms)} & \textbf{CACHED (ms)} & \textbf{Notes} \\
\midrule
Runs ($n$) & 30 & 30 & 3 warm up runs excluded \\
Workload (tx / events / blocks) & \multicolumn{2}{c}{10 / 15 / 10} & One coffee batch query \\
\addlinespace
End to end AQL p50 & 1018.21 & 0.16 & Median reconstruction time \\
End to end AQL mean & 1052.67 & 0.17 & Mean across runs \\
End to end AQL p95 & 1209.67 & 0.19 & Tail latency under evaluated workload \\
Observed end to end max & 1300.85 & 0.98 & Worst observed under test conditions \\
\addlinespace
$T_{\text{receipts}}$ mean & 530.08 & 0.02 & Receipt fetch dominates UNCACHED \\
$T_{\text{timestamps}}$ mean & 522.37 & 0.03 & Block timestamp fetch dominates UNCACHED \\
$T_{\text{decode}}$ mean & 0.15 & 0.11 & CPU overhead negligible \\
$T_{\text{sort}}$ mean & 0.02 & 0.00 & Canonical ordering negligible \\
\bottomrule
\end{tabular}
\end{table}

\noindent \textbf{Interpretation.} Under the evaluated workload, uncached reconstruction is dominated by RPC round trip costs for receipts and block headers, yielding $\mathrm{AQL}_{p95}\approx 1.21$~s. Cached reconstruction demonstrates that compute overhead is negligible once immutable chain data is ingested and cached. This matters for TAFES because lower reconstruction latency reduces the operational cost of audits (increasing feasible audit frequency) and supports accountability mechanisms that rely on timely dispute reconstruction.

\subsection{Model based stress analysis I: batching policy and user perceived anchoring delay}
\label{sec:stress_batching}

\noindent The empirical benchmark in Section~\ref{sec:results_rq3} characterises throughput and inclusion latency, but user experience depends on how the backend batches events before submission. To translate measured anchoring capacity into user perceived delay, we analyse a common batching policy used in service mediated systems. The batcher flushes when it has accumulated $B$ CID commitments, or when a maximum waiting time $\tau$ has elapsed.

\noindent Let events arrive at rate $\lambda$ (events/s). Under this policy, the maximum batching wait for an event is:
\begin{equation}
\label{eq:batch_wait_max}
W_{\max} = \min\left(\tau, \frac{B}{\lambda}\right)
\end{equation}

\noindent and the mean batching wait is approximately $W_{\max}/2$ (events are uniformly distributed within the fill and flush window). The expected anchoring delay is therefore:
\begin{equation}
\label{eq:anchor_delay_batching}
\mathbb{E}[D_{\text{anchor}}] \approx \mathbb{E}[W_{\text{batch}}] + \mathbb{E}[S_{\text{include}}]
\end{equation}
\noindent where $S_{\text{include}}$ is the measured inclusion latency (Table~\ref{tab:op_bench}). This decomposition makes explicit that, under low arrival rates, delay is dominated by the batching window and inclusion latency rather than by throughput limits.

\begin{table}[ht]
\centering
\caption{Anchoring delay under a time and size based batching policy (model based; parameterised by empirical inclusion latency)}
\label{tab:batching_delay}
\setlength{\tabcolsep}{2pt}\scriptsize\renewcommand{\arraystretch}{0.98}
\begin{tabular}{@{}p{0.16\textwidth}p{0.14\textwidth}p{0.14\textwidth}p{0.15\textwidth}p{0.14\textwidth}p{0.15\textwidth}@{}}
\toprule
\textbf{Arrival rate $\lambda$} & \textbf{$W_{\max}$ (s)} & \textbf{$\mathbb{E}[W]$ (s)} & \textbf{$W_{p95}$ (s)} & \textbf{$\mathbb{E}[D]$ (s)} & \textbf{$D_{p95}$ (s)} \\
\midrule
1 events/s & 1.00 & 0.50 & 0.95 & 2.50 & 2.95 \\
10 events/s & 1.00 & 0.50 & 0.95 & 2.50 & 2.95 \\
50 events/s & 1.00 & 0.50 & 0.95 & 2.50 & 2.95 \\
200 events/s & 1.00 & 0.50 & 0.95 & 2.50 & 2.95 \\
600 events/s & 0.85 & 0.43 & 0.81 & 2.43 & 2.81 \\
1200 events/s & 0.43 & 0.21 & 0.41 & 2.21 & 2.41 \\
\bottomrule
\end{tabular}
\end{table}

\noindent \textbf{Interpretation.} For low to moderate workloads, anchoring delay is driven primarily by the chosen maximum batching window plus inclusion latency. Because the measured peak anchoring throughput is high (Table~\ref{tab:op_bench}), the proof of concept has substantial headroom for realistic supply chain workloads. This supports RQ3 and the TAFES Fairness objective because low latency can be achieved without requiring stakeholders to operate specialised infrastructure.

\subsection{Model based stress analysis II: fee pressure and a fairness break even check}
\label{sec:stress_fees_fairness}

\noindent Table~\ref{tab:cost_sensitivity} reports fee sensitivity in absolute terms, but fairness concerns whether traceability overhead becomes material relative to the economic flows that fair trade systems aim to protect. A simple break even check can be constructed using published premium structures. For example, for coffee, a premium is typically defined per unit mass of certified product (e.g., USD per pound) \citep{fairtradeMinPricesPremiums}.

\noindent Let $\Pi$ denote the premium flow per pound (USD/lb). Let $M$ be the batch mass (lb) represented by one proof of concept provenance batch. Then premium flow per batch is $\Pi \cdot M$. A conservative fairness constraint is that anchoring cost should remain below a small fraction $\alpha$ of premium flow (e.g., $\alpha = 1\%$):
\begin{equation}
\label{eq:fairness_constraint}
C_{\text{batch}} \le \alpha \cdot \Pi \cdot M
\end{equation}

\noindent Solving for the minimum mass per batch needed to satisfy this constraint yields:
\begin{equation}
\label{eq:mass_threshold}
M \ge \frac{C_{\text{batch}}}{\alpha \Pi}
\end{equation}

\noindent Equation~\ref{eq:mass_threshold} clarifies the design lever. Anchoring at the lot or batch level, rather than per retail unit, dramatically reduces per unit overhead and helps ensure traceability does not erode the economic intent of fair trade premiums.

\subsection{Model based stress analysis III: evidence availability under churn and replication}
\label{sec:stress_churn}

\noindent The evidence experiment in Table~\ref{tab:evidence_results} reflects a pinned, provider backed configuration. In practice, evidence availability can degrade due to churn (nodes go offline), unpinning, gateway rate limits, or intermittent connectivity. In content addressed systems, integrity remains strong (CID mismatch detects tampering), but availability becomes an operational resilience problem \citep{benet2014ipfs}.

\noindent A lightweight availability model captures the main lever. Suppose an evidence object is redundantly pinned across $k$ independent providers (or nodes), and each provider is available with probability $p$ during a retrieval attempt. Then:
\begin{equation}
\label{eq:availability}
P(\text{retrievable}) = 1 - (1-p)^k
\end{equation}

\begin{table}[ht]
\centering
\caption{Evidence availability under churn with replication and fallback (model based; parameterised by measured fetch latency scale)}
\label{tab:availability_table}
\setlength{\tabcolsep}{2pt}\scriptsize\renewcommand{\arraystretch}{0.98}
\begin{tabular}{@{}p{0.12\textwidth}p{0.08\textwidth}p{0.18\textwidth}p{0.18\textwidth}p{0.38\textwidth}@{}}
\toprule
\textbf{$p$} & \textbf{$k$} & \textbf{$P(\text{retrievable})$} & \textbf{Expected tries} & \textbf{Interpretation} \\
\midrule
0.95 & 1 & 0.9500 & 1.00 & Single provider concentrates availability risk \\
0.95 & 2 & 0.9975 & 1.05 & Two independent pins eliminate most churn related failures \\
0.95 & 3 & 0.9999 & 1.05 & Additional replication yields diminishing returns \\
\addlinespace
0.98 & 1 & 0.9800 & 1.00 & High availability single provider \\
0.98 & 2 & 0.9996 & 1.02 & Near ``four nines'' availability with two replicas \\
0.98 & 3 & 0.999992 & 1.02 & Three replicas yields six nines scale under independence \\
\addlinespace
0.99 & 1 & 0.9900 & 1.00 & Baseline cloud like availability \\
0.99 & 2 & 0.9999 & 1.01 & Two replicas provide very strong availability \\
0.99 & 3 & 0.999999 & 1.01 & Three replicas makes unavailability extremely unlikely \\
\bottomrule
\end{tabular}
\end{table}

\noindent \textbf{Interpretation.} The churn model provides an explicit Safety design lever. Modest replication (e.g., $k=2$) sharply increases availability without weakening integrity. Operationally, this supports a governance requirement: evidence pinning should be treated as a managed control (diverse pinning providers, periodic audits of pin status, and multi gateway retrieval fallbacks).

\subsection{Model based stress analysis IV: oracle risk, audit sampling, and detectability}
\label{sec:stress_oracle_privacy}

\noindent The oracle problem remains a primary risk in certification settings. Actors can submit incorrect data that becomes immutable after anchoring \citep{caldarelli2020oracle,hassan2023oracle}. The proof of concept addresses this risk through attributable submissions, validator attestations, and evidence linkage. A key question for Accountability is whether false inputs are detectable at reasonable operational cost.

\noindent A minimal detection model considers two mechanisms: front door verification at submission time (evidence required; validator rejects missing or invalid evidence), and back door detection via audits (sampling and reconstruction). Let $v$ be the probability that a false event is rejected at submission time, and let $s$ be the audit sampling rate applied to the remaining events. Under independence, a conservative detection probability is:
\begin{equation}
\label{eq:detection_prob}
d = 1 - (1-v)(1-s) = v + (1-v)s
\end{equation}

\begin{table}[ht]
\centering
\caption{Oracle risk detection sensitivity under validator gating and audit sampling (model based)}
\label{tab:oracle_sensitivity}
\setlength{\tabcolsep}{2pt}\scriptsize\renewcommand{\arraystretch}{0.98}
\begin{tabular}{@{}p{0.22\textwidth}p{0.17\textwidth}p{0.17\textwidth}p{0.36\textwidth}@{}}
\toprule
\textbf{Assumptions} & \textbf{Sampling $s$} & \textbf{Detection $d$} & \textbf{Interpretation} \\
\midrule
Moderate front door gating ($v=0.2$) & 1\% & 0.208 & Low audit rates add limited detectability if gating is weak \\
Moderate front door gating ($v=0.2$) & 10\% & 0.280 & Higher sampling materially increases detection for weak gating settings \\
\addlinespace
Strong front door gating ($v=0.6$) & 1\% & 0.604 & Detection dominated by submission time controls \\
Strong front door gating ($v=0.6$) & 10\% & 0.640 & Sampling adds incremental detectability and deterrence \\
\bottomrule
\end{tabular}
\end{table}

\noindent \textbf{Interpretation.} The model suggests two practical implications consistent with TAFES. Reducing actor burden (Fairness) should be paired with enforceable evidence gates (Accountability) so that $v$ is not close to zero. In addition, fast audit reconstruction (Table~\ref{tab:audit_query_latency_strong_conservative}) reduces the marginal cost of increasing $s$, which increases deterrence and supports completeness incentives (Section~\ref{sec:results_completeness}).

\subsubsection{Privacy leakage check (Safety)}
\noindent Safety in this context includes privacy by design. The proof of concept minimises on chain payloads: it anchors lifecycle metadata and evidence commitments (CIDs or hashes) rather than storing evidence contents on chain. This reduces the risk of publishing sensitive producer or commercial data to an immutable ledger. Remaining privacy risks are primarily metadata linkage (e.g., correlating addresses, timing, and repeated identifiers). Operational mitigations include role separated addresses, minimised and hashed identifiers where appropriate, and selective disclosure in the application layer, consistent with the hybrid design rationale in Section~\ref{sec:poc-instantiation} and with supply-chain blockchain work that treats confidentiality as a first-order design constraint \citep{lautenschlager2025strikingBalance}.

\subsection{Consolidated scorecard: TAFES aligned evaluation outcomes}
\label{sec:tafes_scorecard_results}

\noindent Table~\ref{tab:tafes_scorecard} consolidates the results into a scorecard aligned with Table~\ref{tab:tafes-req}. Empirical outcomes are reported where directly measured; stress analysis outcomes are explicitly labelled as model based.

\begin{table}[ht]
\centering
\caption{TAFES scorecard linking principles to measurable outcomes (empirical plus model based)}
\label{tab:tafes_scorecard}
\setlength{\tabcolsep}{2pt}\scriptsize\renewcommand{\arraystretch}{0.98}
\begin{tabular}{@{}p{0.17\textwidth}p{0.41\textwidth}p{0.34\textwidth}@{}}
\toprule
\textbf{TAFES principle} & \textbf{Operational check} & \textbf{Evidence / result summary} \\
\midrule
Transparency &
Complete provenance chain and verifiable evidence linkage &
End to end scenario records all required steps for the demonstrated batch ($C=1.0$ in Eq.~\ref{eq:core_metrics}). Evidence verifiability (empirical): $R=1.00$ and $M=1.00$ across $n=40$ objects (Table~\ref{tab:evidence_results}), yielding $V=1.00$. \\
\addlinespace
Accountability &
Attributable actions; unauthorised actions rejected; audit reconstruction feasible &
Role and status controls and replay protection prevent unauthorised writes and duplicate step anchors (implementation details in Appendix~\ref{app:smart_contracts}). Audit reconstruction performance (empirical): $\mathrm{AQL}_{p95}\approx 1.21$~s under the evaluated workload (Table~\ref{tab:audit_query_latency_strong_conservative}). \\
\addlinespace
Fairness &
Low operational burden; feasible near real time recording &
Service mediated interaction avoids node operation for participants. Anchoring delay under batching (model based): dominated by $\tau$ and inclusion latency under plausible workloads (Table~\ref{tab:batching_delay}). \\
\addlinespace
Ethics &
Supports fair trade outcomes beyond documentation &
The proof of concept provides auditable provenance and evidence linkage that can support premium accountability narratives; deployment scale socio economic impact requires stakeholder pilots and governance evaluation beyond proof of concept scope. \\
\addlinespace
Safety &
Cost viability; privacy by design; resilience under churn &
On chain gas profile supports predictable costs (Table~\ref{tab:op_bench}) with fee sensitivity (Table~\ref{tab:cost_sensitivity}). Evidence availability under churn improves rapidly with replication (model based; Table~\ref{tab:availability_table}). Hybrid design minimises on chain sensitive payloads; residual metadata linkage risk is handled through operational controls. \\
\bottomrule
\end{tabular}
\end{table}

\noindent In summary, the proof of concept provides empirical evidence for infrastructure feasibility and verifiable evidence linkage, and the stress analyses make explicit the operational levers (batching, replication, audit sampling) needed for the design to remain trustworthy under more demanding deployment conditions.

% ==========================================================
\section{Discussion}
\label{sec:discussion}

\subsection{Synthesis of Sections 1--5 (problem, design logic, artefact, and evidence)}
\noindent Sections~\ref{sec:introduction} and \ref{sec:blockchain} position ethical labels as credence claims that are vulnerable to verification gaps, fragmented evidence, and stakeholder mistrust in multi tier supply chains. The analysis highlights that these are not only data integrity problems but also coordination and governance problems, where the locus of trust shifts between certifiers, brands, intermediaries, and consumers \citep{bernards2022veil,balzarova2020conundrum,beck2018governance,fani2025cultivatingTrust,feulner2025intermediaryRoles}. This motivates a design objective that is broader than ``more transparency'': the platform must make verification communicable at the point of purchase, auditable for oversight actors, and operationally feasible for producers with heterogeneous digital capability \citep{sodamin2022fair,kshetri2021developing,ma2025consumerQuality}.

\noindent Section~\ref{sec:design-adr} responds by translating TAFES into implementable requirements and evaluation checks. Conceptually, this operationalisation addresses a recurring critique in the blockchain for sustainability literature: blockchain affordances (immutability, transparency, decentralisation) do not automatically translate into improved sustainability outcomes unless governance, incentives, and process models are designed explicitly \citep{bager2022not,friedman2022sustainability,frizzobarker2020disruptive,bons2020potential,ostern2020blockchainIS}. Methodologically, ADR is appropriate because label authentication is a socio technical problem with evolving stakeholder constraints. Iterative build--intervene--evaluate cycles provide a structured way to align technical mechanisms with stakeholder acceptance and institutional requirements \citep{sein2011adr}.

\noindent Section~\ref{sec:architecture} implements these requirements through an evidence centric, hybrid architecture. Compact integrity commitments (CIDs or hashes and minimal metadata) are recorded on a scalable Layer 2 ledger, while evidence payloads remain off chain in a content addressed store. This design directly targets the adoption constraints identified in Section~\ref{sec:blockchain}, especially scalability and cost and the tension between privacy and regulatory duties, without relying on stakeholders to operate blockchain infrastructure. The architecture also reflects broader research that treats blockchain platforms as governance sensitive market infrastructures, where role definitions, permissions, and exception pathways must be designed alongside the ledger \citep{cao2021governance,beck2018governance,mendling2018bpm,bendig2025multisidedPlatforms,feulner2025intermediaryRoles}.

\noindent Section~\ref{sec:results} provides empirical and model based evidence that the proof of concept can sustain near real time anchoring at low marginal cost when provenance is represented as compact commitments and batched submissions. Equally important, the results clarify that user perceived verification latency is primarily influenced by batching policy and off chain evidence retrieval, not only by ledger throughput. This shifts attention from whether the blockchain can scale to whether the socio technical system can sustain complete, high quality, and retrievable evidence under realistic operational conditions \citep{casino2019systematic,caldarelli2020oracle,alt2025dlt}.

\subsection{Implications for research (TAFES as an operational bridge)}
\noindent The main theoretical contribution is not simply another demonstration that blockchain can support traceability, but an explicit bridge from ethical design principles to implementable platform controls. TAFES treats fairness and safety as first order design objects rather than residual constraints, which aligns with adoption studies showing that capability asymmetries and compliance burdens can undermine participation in sustainable supply chain initiatives \citep{kshetri2021developing,nayal2023antecedents,fani2025cultivatingTrust}. In addition, the proof of concept demonstrates that certification claims that do not rely on trust in a single intermediary must still be grounded in practical governance: role based authority, auditable evidence, and mechanisms for contestation and remediation \citep{bernards2022veil,beck2018governance,feulner2025intermediaryRoles}.

\noindent The paper also contributes to process aware blockchain design by mapping certification and custody workflows into an event model that remains verifiable while avoiding sensitive on chain payloads. This connects to process management perspectives that view smart contracts as a way to enforce and log state transitions, but only if exceptions, permissions, and organisational handoffs are modelled explicitly \citep{mendling2018bpm}. It also extends \textit{Electronic Markets} work on blockchain value chains by showing how stakeholder responsibilities can be translated into a concrete event-and-evidence model for label certification \citep{witt2023valueChains}. As a result, TAFES becomes analytically useful. It can be used to classify design choices, define evaluation criteria, and make trade offs visible (e.g., transparency versus privacy, security versus accessibility) in a way that is directly testable in prototypes and pilots.

\subsection{Implications for practice and policy (what to do differently)}
\noindent From a managerial perspective, the results suggest three practical design levers. First, batch level anchoring is a necessary economic control. Recording provenance at the lot or batch level, rather than per retail unit, maintains verifiability while keeping traceability overhead negligible relative to product value and premiums. Second, evidence operations are operationally decisive. Organisations should treat evidence pinning and replication, retrieval fallback, and evidence quality controls as core operational responsibilities, not as peripheral storage choices. Third, adoption requires interface and governance design, not only ledger selection. Producers, retailers, and certifiers must be able to participate without specialised blockchain operations, and regulators must be able to reconstruct audits efficiently from durable logs \citep{cao2021governance,kouhizadeh2021adoption,mafike2026interoperability}.

\noindent For policy actors and certification bodies, the proof of concept supports an integration strategy rather than a replacement strategy. Certification organisations can use the platform as a complementary audit and evidence layer. Anchoring certificates and inspection evidence provides tamper evident traceability while preserving existing standards, audit routines, and legal accountability structures. This framing can also reduce institutional resistance, which prior research identifies as a common barrier when blockchain is presented as a full substitution for established governance arrangements \citep{friedman2022sustainability,bager2022not,feulner2025intermediaryRoles}. Finally, the platform's minimal on chain data design is consistent with privacy by design expectations. It limits exposure while still enabling verification, shifting compliance focus toward access control, consent, and metadata management in the application layer.

\subsection{Limitations}
\noindent The study has limitations typical of early stage design science artefacts. The proof of concept and performance measurements are conducted in a bounded environment (test network and provider backed evidence services), and the evaluation prioritises feasibility, integrity, and verifiability over long run behavioural outcomes. In particular, the paper does not claim that anchoring alone improves producer welfare or eliminates greenwashing; those outcomes depend on governance, incentives, audit enforcement, and market adoption \citep{bager2022not,bernards2022veil}. Moreover, although the architecture minimises sensitive on chain data, metadata linkage and organisational data practices remain potential privacy risks that must be managed with operational controls and governance arrangements.

\subsection{Ongoing and future work (research and development agenda)}
\noindent Building on the findings in Section~\ref{sec:results} and the adoption barriers in Section~\ref{sec:blockchain}, ongoing work is prioritised in five areas.

\begin{enumerate}
  \item \textbf{Integration with certification regimes and standards.} The platform should be evaluated in joint pilots with existing certifiers to map standard operating procedures into verifiable event and evidence templates, and to test how anchored evidence supports audit cycles, dispute handling, and claims substantiation at scale \citep{santos2021thirdparty,kshetri2021developing}.
  \item \textbf{Data quality controls and oracle mitigation.} Future iterations should combine evidence gating at submission, structured audit sampling, and selective sensor or third party attestations for high risk steps (e.g., shipment conditions, processing claims). This directly targets the ``garbage in, garbage out'' limitation of ledger integrity \citep{caldarelli2020oracle,hassan2023oracle,hasan2024smartagriculture}.
  \item \textbf{Privacy preserving verification.} The hybrid design provides a baseline; next steps include selective disclosure and privacy preserving proof techniques for validating compliance attributes without revealing sensitive commercial or personal data. This would reduce metadata leakage risk while retaining auditability under TAFES Safety \citep{lautenschlager2025strikingBalance}.
  \item \textbf{Interoperability and platform governance.} To avoid recreating digital silos, future work should adopt interoperable identifiers and process models that can connect to retailer systems, certification databases, and, where relevant, cross chain networks. Governance design should be treated as a first order concern, with explicit policies for actor onboarding, revocation, and accountability under multi stakeholder conditions \citep{cao2021governance,beck2018governance,casino2019systematic,mafike2026interoperability,feulner2025intermediaryRoles}.
  \item \textbf{Impact and adoption evaluation.} Beyond technical KPIs, field studies should measure consumer comprehension and trust effects of mediated verification experiences, producer participation costs and fairness outcomes, and organisational adoption drivers such as readiness, stakeholder pressure, and analytics capability \citep{sodamin2022fair,nayal2023antecedents,khan2025greendata,ma2025consumerQuality,fani2025cultivatingTrust}.
\end{enumerate}

\noindent In sum, the proof of concept establishes infrastructure level feasibility and a replicable design logic, but responsible scaling requires systematic governance integration, evidence operations, and stakeholder facing evaluation in real deployment contexts.

% ==========================================================
\section{Conclusion}
\label{sec:conclusion}

\noindent This study developed and evaluated a TAFES aligned digital platform for authenticating fair trade and related labels using a hybrid Ethereum Layer 2 and off chain evidence architecture. By translating TAFES into implementable requirements and checks and executing the work through ADR, the paper moves beyond generic claims about blockchain transparency toward a concrete, testable blueprint for evidence based label verification in multi stakeholder supply chains.

\noindent The proof of concept demonstrates that anchoring compact, content addressed evidence commitments can support low cost, near real time provenance logging while maintaining auditability and reducing reliance on opaque, document centric verification. The results also show that end to end verification performance depends as much on batching policy and evidence operations as on ledger throughput, reinforcing the need to treat governance, usability, and operational resilience as co equal design objects.

\noindent Future work should prioritise certification body integration, oracle risk mitigation, privacy preserving verification, and rigorous field evaluation of adoption and impact. These steps are necessary to validate that platform level transparency translates into credible claims, accountable governance, and fair participation in real market conditions.

\backmatter

\section*{Acknowledgments}
The authors have no acknowledgments to declare.

\section*{Disclosure statement}
No potential conflict of interest is reported by the authors.

\section*{Code availability}
All code, configuration files, and scripts required to reproduce the empirical benchmarks and model based stress analyses reported in Section~\ref{sec:results} are publicly available at: \url{https://github.com/P-HOW/fairtrade-infra-benchmarks}.

% ==========================================================

\begin{appendices}

\section{On chain contract suite and interface summary}
\label{app:smart_contracts}

\noindent This appendix documents implementation specific contract components and interface elements referenced at a high level in the main text. The purpose is to keep Section~\ref{sec:results} focused on measurable outcomes while maintaining full technical traceability for replication.

\subsection{Contract components}
\noindent The proof of concept uses a modular smart contract suite consisting of the following components:
\begin{itemize}
    \item \texttt{ActorRegistry}: manages actor registration, role assignment, and status (e.g., active, suspended, revoked). This component underpins authorisation checks used across write operations.
    \item \texttt{CidRollup}: anchors batched CID commitments for lifecycle events and evidence pointers. This component is the primary on chain workload in the anchoring throughput benchmarks.
    \item \texttt{DocumentRegistry}: associates document identifiers with CID commitments and emits indexed events to support audit reconstruction.
    \item \texttt{ProcessManager}: encodes lifecycle state transitions and enforces valid step progression for a product or batch identifier.
    \item \texttt{PaymentRouter}: provides an optional routing layer for conditional payments or premium related settlement logic (not required for the anchoring throughput results reported in Section~\ref{sec:results_rq3}).
\end{itemize}

\subsection{Batch anchoring call and measured gas profile}
\noindent The anchoring benchmark in Section~\ref{sec:results_rq3} exercises the batched anchoring entry point \texttt{submitCidBatch}, which appends a batch of CID commitments within a single transaction. The benchmark records the gas used per transaction and reports mean gas per CID commitment by dividing the observed transaction gas by the number of commitments in the batch (Table~\ref{tab:op_bench}).

\subsection{Replay protection keying}
\noindent The accountability evaluation in Section~\ref{sec:results_accountability} relies on deterministic replay protection for step anchoring. The implementation maintains a replay guard keyed by a pair of identifiers, \texttt{(productId, stepId)}, stored as \texttt{usedStepKey}. Once an anchor is accepted for a given pair, subsequent attempts for the same pair are rejected.

\subsection{Authorisation checks and actor status}
\noindent Write operations that submit lifecycle anchors require the caller to be registered and in an active status. Calls by unregistered accounts, or by accounts marked suspended or revoked, revert deterministically. These checks are enforced at the contract level so they remain effective even if an off chain service layer is misconfigured.

\section{Repository and harness entry points for reproducibility}
\label{app:reproducibility}

\noindent This appendix lists the concrete entry points and implementation details that are intentionally omitted from Section~\ref{sec:results} to preserve an academic presentation style while ensuring full reproducibility.

\subsection{Evaluation harness environment}
\noindent The evaluation harness is implemented in TypeScript and executed in a Node.js environment (Node.js \texttt{v22.21.1}). Smart contract interaction is managed through a Hardhat 3 project. The harness uses an HTTP client for JSON RPC calls to OP Sepolia endpoints and uses a deterministic CID hashing implementation to recompute and verify content identifiers.

\subsection{Evidence verifiability harness}
\noindent The IPFS evidence loop experiment reported in Section~\ref{sec:results_evidence} is implemented in the repository under:
\begin{quote}
\texttt{scripts/sim-filebase-evidence.ts}
\end{quote}
\noindent The harness generates test objects across four file sizes, computes the expected CID, uploads the object to the configured IPFS compatible RPC endpoint with pinning enabled, retrieves the object by CID, and recomputes the CID from retrieved bytes to verify integrity.

\subsection{RPC endpoints and IPFS RPC operations}
\noindent The evaluation uses OP Sepolia JSON RPC endpoints (public and private) configured through environment variables. Evidence storage uses an IPFS compatible pinning provider (Filebase) accessed via standard IPFS RPC operations for adding and retrieving content. Concretely, the harness uses the IPFS RPC API endpoints \texttt{/api/v0/add} (with pinning enabled and CID version set to 1) and \texttt{/api/v0/cat} for retrieval.

\subsection{Anchoring benchmark harness}
\noindent The anchoring throughput and batch size measurements in Section~\ref{sec:results_rq3} are produced by repository scripts that submit repeated batched anchoring transactions and record inclusion delay and gas usage. The harness supports multi account submission and multi provider comparisons to probe rate limiting and throughput. The exact script names and invocation commands are documented in the repository's README, along with configuration templates for RPC URLs and funded test accounts.

\subsection{Audit query latency harness}
\noindent The AQL measurements in Table~\ref{tab:audit_query_latency_strong_conservative} are produced by a harness that reconstructs an ordered audit trail for a single batch by fetching transaction receipts and block headers, decoding indexed event logs, ordering by block number and log index, and attaching timestamps. The harness reports both uncached and cached regimes and is documented in the repository with reproducible run commands and parameters.

\end{appendices}

\bibliography{sn-bibliography}

@article{abdu2021wtpcoffee,
  author  = {Abdu, Nizam and Mutuku, Judith},
  title   = {Willingness to pay for socially responsible products: A meta-analysis of coffee ecolabelling},
  journal = {Heliyon},
  year    = {2021},
  volume  = {7},
  number  = {6},
  pages   = {e07043},
  doi     = {10.1016/j.heliyon.2021.e07043}
}

@article{consumerWTPsustainableCoffee2024,
  author  = {Santos, Jos{\'e} and Raggi, Meri and Viaggi, Davide},
  title   = {Consumer Willingness-to-Pay for Sustainable Coffee: Evidence from Spain},
  journal = {Sustainability},
  year    = {2024},
  volume  = {16},
  number  = {8},
  pages   = {3222},
  doi     = {10.3390/su16083222}
}

@article{foodCredenceAttributes2023,
  author  = {Schrobback, Patrick and Zhang, Amy and Loechel, Bernd and Ricketts, Kristin and Ingham, Amy},
  title   = {Food Credence Attributes: A Conceptual Framework of Supply Chain Stakeholders, Their Motives, and Mechanisms to Address Information Asymmetry},
  journal = {Foods},
  year    = {2023},
  volume  = {12},
  number  = {3},
  pages   = {538},
  doi     = {10.3390/foods12030538}
}

@inproceedings{langer2007ecolabelconfusion,
  author    = {Langer, Alexandra and Eisend, Martin and Ku{\ss}, Alfred},
  title     = {The Impact of Eco-Labels on Consumers: Less Information, More Confusion?},
  booktitle = {European Advances in Consumer Research},
  year      = {2007},
  volume    = {8},
  pages     = {338--339},
  publisher = {Association for Consumer Research}
}

@article{sodamin2022fair,
  author  = {Sodamin, D. and Vanek, J. and Ulman, M. and Simek, P.},
  title   = {Fair label versus blockchain technology from the consumer perspective: Towards a comprehensive research agenda},
  journal = {AGRIS On-line Papers in Economics and Informatics},
  year    = {2022},
  volume  = {14},
  number  = {2},
  pages   = {95--108},
  doi     = {10.7160/aol.2022.140209}
}

@article{lou2024denim,
  author  = {Lou, X. and Xu, Y.},
  title   = {Consumption of sustainable denim products: The contribution of blockchain certified eco-labels},
  journal = {Journal of Theoretical and Applied Electronic Commerce Research},
  year    = {2024},
  volume  = {19},
  number  = {1},
  pages   = {396--411},
  doi     = {10.3390/jtaer19010021}
}

@article{hilten2020blockchain,
  author  = {Hilten, M. and Ongena, G. and Ravesteijn, P.},
  title   = {Blockchain for organic food traceability: Case studies on drivers and challenges},
  journal = {Frontiers in Blockchain},
  year    = {2020},
  volume  = {3},
  pages   = {567175},
  doi     = {10.3389/fbloc.2020.567175}
}

@article{katsikouli2020benefits,
  author  = {Katsikouli, P. and Wilde, A. and Dragoni, N. and H{\o}gh-Jensen, H.},
  title   = {On the benefits and challenges of blockchains for managing food supply chains},
  journal = {Journal of the Science of Food and Agriculture},
  year    = {2020},
  volume  = {101},
  number  = {6},
  pages   = {2175--2181},
  doi     = {10.1002/jsfa.10883}
}

@article{bernards2022veil,
  author  = {Bernards, N. and Campbell-Verduyn, M. and Rodima-Taylor, D.},
  title   = {The veil of transparency: Blockchain and sustainability governance in global supply chains},
  journal = {Environment and Planning C: Politics and Space},
  year    = {2022},
  volume  = {42},
  number  = {4},
  pages   = {742--760},
  doi     = {10.1177/23996544221142763}
}

@article{balzarova2020conundrum,
  author  = {Balzarova, M. and Cohen, D.},
  title   = {The blockchain technology conundrum: Quis custodiet ipsos custodes?},
  journal = {Current Opinion in Environmental Sustainability},
  year    = {2020},
  volume  = {45},
  pages   = {42--48},
  doi     = {10.1016/j.cosust.2020.08.016}
}

@article{friedman2022sustainability,
  author  = {Friedman, N. and Ormiston, J.},
  title   = {Blockchain as a sustainability-oriented innovation? Opportunities for and resistance to blockchain technology as a driver of sustainability in global food supply chains},
  journal = {Technological Forecasting and Social Change},
  year    = {2022},
  volume  = {175},
  pages   = {121403},
  doi     = {10.1016/j.techfore.2021.121403}
}

@article{fani2025cultivatingTrust,
  author  = {Fani, Virginia and Ciccullo, Federica and Bandinelli, Romeo and Pero, Margherita},
  title   = {Cultivating trust: An empirical exploration of blockchain's adoption within the Italian wine supply chain},
  journal = {Electronic Markets},
  year    = {2025},
  volume  = {35},
  number  = {1},
  pages   = {35},
  doi     = {10.1007/s12525-025-00782-y},
  url     = {https://link.springer.com/article/10.1007/s12525-025-00782-y}
}

@article{bager2022not,
  author  = {Bager, S. and Singh, C. and Persson, U.},
  title   = {Blockchain is not a silver bullet for agro-food supply chain sustainability: Insights from a coffee case study},
  journal = {Current Research in Environmental Sustainability},
  year    = {2022},
  volume  = {4},
  pages   = {100163},
  doi     = {10.1016/j.crsust.2022.100163}
}

@article{kshetri2021developing,
  author  = {Kshetri, Nir},
  title   = {Blockchain and sustainable supply chain management in developing countries},
  journal = {International Journal of Information Management},
  year    = {2021},
  volume  = {60},
  pages   = {102376},
  doi     = {10.1016/j.ijinfomgt.2021.102376}
}

@article{kouhizadeh2021adoption,
  author  = {Kouhizadeh, Mahtab and Saberi, Sara and Sarkis, Joseph},
  title   = {Blockchain technology and the sustainable supply chain: Theoretically exploring adoption barriers},
  journal = {International Journal of Production Economics},
  year    = {2021},
  volume  = {231},
  pages   = {107831},
  doi     = {10.1016/j.ijpe.2020.107831}
}

@article{park2021effect,
  author  = {Park, A. and Li, H.},
  title   = {The effect of blockchain technology on supply chain sustainability performances},
  journal = {Sustainability},
  year    = {2021},
  volume  = {13},
  number  = {4},
  pages   = {1726},
  doi     = {10.3390/su13041726}
}

@article{alt2025dlt,
  author  = {Alt, Rainer and Gr{\"a}ser, Max},
  title   = {Distributed ledger technology},
  journal = {Electronic Markets},
  year    = {2025},
  volume  = {35},
  number  = {1},
  pages   = {53},
  doi     = {10.1007/s12525-025-00784-w},
  url     = {https://link.springer.com/article/10.1007/s12525-025-00784-w}
}

@article{bons2020potential,
  author  = {Bons, Roger W. H. and Versendaal, Johan and Zavolokina, Liudmila and Shi, Weidong Larry},
  title   = {Potential and limits of Blockchain technology for networked businesses},
  journal = {Electronic Markets},
  year    = {2020},
  volume  = {30},
  number  = {2},
  pages   = {189--194},
  doi     = {10.1007/s12525-020-00421-8},
  url     = {https://link.springer.com/article/10.1007/s12525-020-00421-8},
  note    = {Preface}
}

@article{ostern2020blockchainIS,
  author  = {Ostern, Nadine Kathrin},
  title   = {Blockchain in the IS research discipline: A discussion of terminology and concepts},
  journal = {Electronic Markets},
  year    = {2020},
  volume  = {30},
  number  = {2},
  pages   = {195--210},
  doi     = {10.1007/s12525-019-00387-2},
  url     = {https://link.springer.com/article/10.1007/s12525-019-00387-2}
}

@article{beck2018governance,
  author  = {Beck, Roman and M{\"u}ller-Bloch, Christoph and King, John Leslie},
  title   = {Governance in the Blockchain Economy: A Framework and Research Agenda},
  journal = {Journal of the Association for Information Systems},
  year    = {2018},
  volume  = {19},
  number  = {10},
  pages   = {1020--1034},
  doi     = {10.17705/1jais.00518}
}

@article{frizzobarker2020disruptive,
  author  = {Frizzo-Barker, Julie and Chow-White, Peter A. and Adams, Paul R. and Mentanko, Jim and Ha, David and Green, Shawn},
  title   = {Blockchain as a Disruptive Technology for Business: A Systematic Review},
  journal = {International Journal of Information Management},
  year    = {2020},
  volume  = {51},
  pages   = {102029},
  doi     = {10.1016/j.ijinfomgt.2019.10.014}
}

@article{casino2019systematic,
  author  = {Casino, Fran and Dasaklis, Thomas K. and Patsakis, Constantinos},
  title   = {A systematic literature review of blockchain-based applications: Current status, classification and open issues},
  journal = {Telematics and Informatics},
  year    = {2019},
  volume  = {36},
  pages   = {55--81},
  doi     = {10.1016/j.tele.2018.11.006}
}

@article{mendling2018bpm,
  author  = {Mendling, Jan and Weber, Ingo and van der Aalst, Wil M. P. and others},
  title   = {Blockchains for Business Process Management -- Challenges and Opportunities},
  journal = {ACM Transactions on Management Information Systems},
  year    = {2018},
  volume  = {9},
  number  = {1},
  pages   = {4:1--4:16},
  doi     = {10.1145/3183367}
}

@article{bendig2025multisidedPlatforms,
  author  = {Bendig, David and Charlet, Maximilian},
  title   = {Opportunities and challenges of blockchain for multi-sided platforms},
  journal = {Electronic Markets},
  year    = {2025},
  volume  = {35},
  number  = {1},
  pages   = {25},
  doi     = {10.1007/s12525-025-00765-z},
  url     = {https://link.springer.com/article/10.1007/s12525-025-00765-z}
}

@article{feulner2025intermediaryRoles,
  author  = {Feulner, Simon and Guggenberger, Tobias and Stoetzer, Jens-Christian and Urbach, Nils},
  title   = {Beyond disintermediation: A multiple case study of emerging intermediary roles in blockchain applications},
  journal = {Electronic Markets},
  year    = {2025},
  volume  = {35},
  number  = {1},
  pages   = {98},
  doi     = {10.1007/s12525-025-00832-5},
  url     = {https://link.springer.com/article/10.1007/s12525-025-00832-5}
}

@article{lautenschlager2025strikingBalance,
  author  = {Lautenschlager, Jonathan and Stramm, Jan and Guggenberger, Tobias and Urbach, Nils},
  title   = {Striking a balance: Designing a blockchain-based solution to navigate coopetition dynamics in supply chain management},
  journal = {Electronic Markets},
  year    = {2025},
  volume  = {35},
  number  = {1},
  pages   = {70},
  doi     = {10.1007/s12525-025-00809-4},
  url     = {https://link.springer.com/article/10.1007/s12525-025-00809-4}
}

@article{mafike2026interoperability,
  author  = {Mafike, Senate Sylvia and Mawela, Tendani},
  title   = {An enterprise framework for blockchain interoperability},
  journal = {Electronic Markets},
  year    = {2026},
  volume  = {36},
  number  = {1},
  pages   = {14},
  doi     = {10.1007/s12525-025-00869-6},
  url     = {https://link.springer.com/article/10.1007/s12525-025-00869-6}
}

@article{sein2011adr,
  author  = {Sein, Maung K. and Henfridsson, Ola and Purao, Sandeep and Rossi, Matti and Lindgren, Rikard},
  title   = {Action Design Research},
  journal = {MIS Quarterly},
  year    = {2011},
  volume  = {35},
  number  = {1},
  pages   = {37--56},
  doi     = {10.2307/23043488}
}

@inproceedings{jensen2019dsr,
  author    = {Jensen, T. and Asheim, A.},
  title     = {The DSR methodology in blockchain research},
  booktitle = {Proceedings of the International Conference on Information Systems (ICIS)},
  year      = {2019},
  pages     = {1--9}
}

@article{hevner2004dsr,
  author  = {Hevner, Alan R. and March, Salvatore T. and Park, Jinsoo and Ram, Sudha},
  title   = {Design Science in Information Systems Research},
  journal = {MIS Quarterly},
  year    = {2004},
  volume  = {28},
  number  = {1},
  pages   = {75--105},
  doi     = {10.2307/25148625}
}

@article{gregor2013dsr,
  author  = {Gregor, Shirley and Hevner, Alan R.},
  title   = {Positioning and Presenting Design Science Research for Maximum Impact},
  journal = {MIS Quarterly},
  year    = {2013},
  volume  = {37},
  number  = {2},
  pages   = {337--355},
  doi     = {10.25300/MISQ/2013/37.2.01}
}

@article{zavolokina2020buyersLemons,
  author  = {Zavolokina, Liudmila and Miscione, Gianluca and Schwabe, Gerhard},
  title   = {Buyers of 'lemons': How can a blockchain platform address buyers' needs in the market for 'lemons'?},
  journal = {Electronic Markets},
  year    = {2020},
  volume  = {30},
  number  = {2},
  pages   = {227--239},
  doi     = {10.1007/s12525-019-00380-9},
  url     = {https://link.springer.com/article/10.1007/s12525-019-00380-9}
}

@article{bauer2020trustedCarData,
  author  = {Bauer, Ingrid and Zavolokina, Liudmila and Schwabe, Gerhard},
  title   = {Is there a market for trusted car data?},
  journal = {Electronic Markets},
  year    = {2020},
  volume  = {30},
  number  = {2},
  pages   = {211--225},
  doi     = {10.1007/s12525-019-00368-5},
  url     = {https://link.springer.com/article/10.1007/s12525-019-00368-5}
}

@article{alt2020blockchainMarkets,
  author  = {Alt, Rainer},
  title   = {Electronic Markets on blockchain markets},
  journal = {Electronic Markets},
  year    = {2020},
  volume  = {30},
  number  = {2},
  pages   = {181--188},
  doi     = {10.1007/s12525-020-00428-1},
  url     = {https://link.springer.com/article/10.1007/s12525-020-00428-1},
  note    = {Editorial}
}

@article{guo2020fashion,
  author  = {Guo, S. and Sun, X. and Lam, H.},
  title   = {Applications of blockchain technology in sustainable fashion supply chains: Operational transparency and environmental efforts},
  journal = {IEEE Transactions on Engineering Management},
  year    = {2020},
  volume  = {69},
  number  = {4},
  pages   = {1535--1551},
  doi     = {10.1109/TEM.2020.3034216}
}

@article{nikolakis2018eve,
  author  = {Nikolakis, W. and John, L. and Krishnan, H.},
  title   = {How blockchain can shape sustainable global value chains: An evidence, verifiability, and enforceability (EVE) framework},
  journal = {Sustainability},
  year    = {2018},
  volume  = {10},
  number  = {11},
  pages   = {3926},
  doi     = {10.3390/su10113926}
}

@article{santos2021thirdparty,
  author  = {Santos, R. and Torrisi, N. and Pantoni, R.},
  title   = {Third party certification of agri-food supply chain using smart contracts and blockchain tokens},
  journal = {Sensors},
  year    = {2021},
  volume  = {21},
  number  = {16},
  pages   = {5307},
  doi     = {10.3390/s21165307}
}

@article{agrawal2021textile,
  author  = {Agrawal, T. and Kumar, V. and Pal, R. and Wang, L. and Chen, Y.},
  title   = {Blockchain-based framework for supply chain traceability: A case example of textile and clothing industry},
  journal = {Computers \& Industrial Engineering},
  year    = {2021},
  volume  = {154},
  pages   = {107130},
  doi     = {10.1016/j.cie.2021.107130}
}

@article{chandan2023sdgs,
  author  = {Chandan, A. and John, M. and Potdar, V.},
  title   = {Achieving UN SDGs in food supply chain using blockchain technology},
  journal = {Sustainability},
  year    = {2023},
  volume  = {15},
  number  = {3},
  pages   = {2109},
  doi     = {10.3390/su15032109}
}

@article{hasan2024smartagriculture,
  author  = {Hasan, H. and Musamih, A. and Salah, K. and Jayaraman, R. and Omar, M. and Arshad, J. and Boscovic, D.},
  title   = {Smart agriculture assurance: IoT and blockchain for trusted sustainable produce},
  journal = {Computers and Electronics in Agriculture},
  year    = {2024},
  volume  = {224},
  pages   = {109184},
  doi     = {10.1016/j.compag.2024.109184}
}

@article{stopfer2024wood,
  author  = {Stopfer, L. and Kaulen, A. and Purf{\"u}rst, T.},
  title   = {Potential of blockchain technology in wood supply chains},
  journal = {Computers and Electronics in Agriculture},
  year    = {2024},
  volume  = {216},
  pages   = {108496},
  doi     = {10.1016/j.compag.2023.108496}
}

@article{liu2023impact,
  author  = {Liu, H. and Zhang, R. and He, G. and Lamrabet, A. and Fu, S.},
  title   = {The impact of blockchain technology on the online purchase behavior of green agricultural products},
  journal = {Journal of Retailing and Consumer Services},
  year    = {2023},
  volume  = {74},
  pages   = {103387},
  doi     = {10.1016/j.jretconser.2023.103387}
}

@article{xiaoyong2024certifying,
  author  = {Xiaoyong, L. and Dai, D.},
  title   = {Certifying greenness: Blockchain's impact on eco-friendly products in a competitive market},
  journal = {IEEE Access},
  year    = {2024},
  volume  = {12},
  pages   = {782--793},
  doi     = {10.1109/ACCESS.2023.3347743}
}

@article{ma2025consumerQuality,
  author  = {Ma, Benedict Jun and Liu, Samuel Shuai and Huang, George Q. and Ng, Chi-To},
  title   = {How does consumer quality preference impact blockchain adoption in supply chains?},
  journal = {Electronic Markets},
  year    = {2025},
  volume  = {35},
  number  = {1},
  pages   = {17},
  doi     = {10.1007/s12525-025-00767-x},
  url     = {https://link.springer.com/article/10.1007/s12525-025-00767-x}
}

@article{sanka2021scalability,
  author  = {Sanka, Anil I. and Cheung, Ryan C. C.},
  title   = {A systematic review of blockchain scalability: issues, solutions, analysis and future research},
  journal = {Journal of Network and Computer Applications},
  year    = {2021},
  volume  = {195},
  pages   = {103232},
  doi     = {10.1016/j.jnca.2021.103232}
}

@misc{kostamis2021ethereumdatastores,
  author        = {Kostamis, Periklis and Sendros, Andreas and Efraimidis, Pavlos},
  title         = {Exploring Ethereum's Data Stores: A Cost and Performance Comparison},
  year          = {2021},
  eprint        = {2105.10520},
  archivePrefix = {arXiv},
  primaryClass  = {cs.DB},
  url           = {https://arxiv.org/abs/2105.10520}
}

@article{caldarelli2020oracle,
  author  = {Caldarelli, Giulio and Rossignoli, Cecilia and Zardini, Alessandro},
  title   = {Overcoming the Blockchain Oracle Problem in the Traceability of Non-Fungible Products},
  journal = {Sustainability},
  year    = {2020},
  volume  = {12},
  number  = {6},
  pages   = {2391},
  doi     = {10.3390/su12062391}
}

@article{hassan2023oracle,
  author  = {Hassan, Ammar and Makhdoom, Imran and Iqbal, Waseem and Ahmad, Awais and Raza, Asad},
  title   = {From trust to truth: Advancements in mitigating the Blockchain Oracle problem},
  journal = {Journal of Network and Computer Applications},
  year    = {2023},
  volume  = {217},
  pages   = {103672},
  doi     = {10.1016/j.jnca.2023.103672}
}

@article{rejeb2020food,
  author  = {Rejeb, A. and Keogh, J. and Zailani, S. and Treiblmaier, H. and Rejeb, K.},
  title   = {Blockchain technology in the food industry: A review of potentials, challenges and future research directions},
  journal = {Logistics},
  year    = {2020},
  volume  = {4},
  number  = {4},
  pages   = {27},
  doi     = {10.3390/logistics4040027}
}

@article{erol2021scrutinizing,
  author  = {Erol, I. and Ar, I. and Peker, I.},
  title   = {Scrutinizing blockchain applicability in sustainable supply chains through an integrated fuzzy multi-criteria decision making framework},
  journal = {Applied Soft Computing},
  year    = {2021},
  volume  = {116},
  pages   = {108331},
  doi     = {10.1016/j.asoc.2021.108331}
}

@article{cao2021governance,
  author  = {Cao, Shoufeng and Miller, Thomas and Foth, Marcus and Powell, Warwick and Boyen, Xavier and Turner-Morris, Charles},
  title   = {Integrating On-chain and Off-chain Governance for Supply Chain Transparency and Integrity},
  journal = {arXiv preprint},
  year    = {2021},
  doi     = {10.48550/arXiv.2111.08455},
  url     = {https://arxiv.org/abs/2111.08455}
}

@article{nayal2023antecedents,
  author  = {Nayal, Kanak and Raut, Rakesh and Narkhede, Bharat and Priyadarshinee, Pragati and Panchal, Gautam and Gedam, Vivek},
  title   = {Antecedents for blockchain technology-enabled sustainable agriculture supply chain},
  journal = {Annals of Operations Research},
  year    = {2023},
  volume  = {327},
  number  = {1},
  pages   = {293--337},
  doi     = {10.1007/s10479-021-04423-3}
}

@article{khan2025greendata,
  author  = {Khan, S. A. R. and Godil, D. I. and Jabbour, C. J. C. and others},
  title   = {Green data analytics, blockchain technology for sustainable development, and sustainable supply chain practices: evidence from small and medium enterprises},
  journal = {Annals of Operations Research},
  year    = {2025},
  volume  = {350},
  pages   = {603--627},
  doi     = {10.1007/s10479-021-04275-x}
}

@incollection{sharma2025tafes,
  author    = {Sharma, R. and Loucif, S. and Khalil, A. and Zahid, A.},
  title     = {A Manifesto for Responsible AI: Healthcare Use-Case of the TAFES Framework (Chapter 28)},
  booktitle = {Information System Design: Big Data Analytics and Data Science -- Proceedings of Ninth International Conference on Information System Design and Intelligent Applications (ISDIA 2025), Volume 3},
  series    = {Lecture Notes in Networks and Systems},
  volume    = {1539},
  year      = {2025},
  publisher = {Springer}
}

@article{jahanbin20213tic,
  author  = {Jahanbin, P. and Sharma, R. and Wingreen, S. and Kshetri, N. and Choo, K. K. R.},
  title   = {Towards CRISP-BC: 3TIC specification framework for blockchain use-cases},
  journal = {IET Blockchain},
  year    = {2021},
  volume  = {1},
  number  = {2},
  pages   = {89--102}
}

@article{peffers2007dsrmethod,
  author  = {Peffers, Ken and Tuunanen, Tuure and Rothenberger, Marcus A. and Chatterjee, Samir},
  title   = {A Design Science Research Methodology for Information Systems Research},
  journal = {Journal of Management Information Systems},
  year    = {2007},
  volume  = {24},
  number  = {3},
  pages   = {45--77},
  doi     = {10.2753/MIS0742-1222240302}
}

@article{dionysis2022tpbcoffee,
  author  = {Dionysis, S. and Chesney, T. and McAuley, D.},
  title   = {Examining the influential factors of consumer purchase intentions for blockchain traceable coffee using the theory of planned behaviour},
  journal = {British Food Journal},
  year    = {2022},
  volume  = {125},
  number  = {7},
  pages   = {2420--2440},
  doi     = {10.1108/bfj-05-2021-0541}
}

@article{contini2023credence,
  author  = {Contini, C. and Boncinelli, F. and Piracci, G. and Scozzafava, G. and Casini, L.},
  title   = {Can blockchain technology strengthen consumer preferences for credence attributes?},
  journal = {Agricultural and Food Economics},
  year    = {2023},
  volume  = {11},
  number  = {1},
  pages   = {1--17},
  doi     = {10.1186/s40100-023-00270-x}
}

@article{owsianowski2025linking,
  author  = {Owsianowski, J. and Bitsch, V.},
  title   = {Linking consumers to producers in fair trade supply chains with the use of blockchain technology},
  journal = {International Journal on Food System Dynamics},
  year    = {2025},
  volume  = {16},
  number  = {1},
  pages   = {1--15},
  doi     = {10.1163/18696945-bja00013}
}

@article{samoggia2025promised,
  author  = {Samoggia, A. and Fantini, A. and Ghelfi, R.},
  title   = {The promised potential of blockchain technology for transparency and fairness in agri-food chains: Insights from the coffee sector},
  journal = {Frontiers in Sustainable Food Systems},
  year    = {2025},
  volume  = {8},
  pages   = {1401735},
  doi     = {10.3389/fsufs.2025.1401735}
}

@misc{benet2014ipfs,
  author        = {Benet, Juan},
  title         = {IPFS -- Content Addressed, Versioned, P2P File System},
  year          = {2014},
  howpublished  = {arXiv preprint},
  eprint        = {1407.3561},
  archivePrefix = {arXiv},
  primaryClass  = {cs.NI},
  url           = {https://arxiv.org/abs/1407.3561}
}

@misc{optimismSepoliaDocs,
  author       = {{Optimism Documentation}},
  title        = {OP Sepolia Testnet},
  year         = {2026},
  howpublished = {\url{https://docs.optimism.io/}},
  note         = {Accessed 25 February 2026}
}

@misc{opStackGlossary,
  author       = {{Optimism Documentation}},
  title        = {OP Stack Glossary},
  year         = {2026},
  howpublished = {\url{https://docs.optimism.io/}},
  note         = {Accessed 25 February 2026}
}

@techreport{fairtradeMinPricesPremiums,
  author      = {{Fairtrade International}},
  title       = {Fairtrade Minimum Price and Premium Table},
  year        = {2026},
  institution = {Fairtrade International},
  url         = {https://files.fairtrade.net/},
  note        = {Accessed 25 February 2026}
}

@article{witt2023valueChains,
  author  = {Witt, Josepha and Schoop, Mareike},
  title   = {Blockchain technology in e-business value chains},
  journal = {Electronic Markets},
  year    = {2023},
  volume  = {33},
  number  = {1},
  pages   = {15},
  doi     = {10.1007/s12525-023-00636-5},
  url     = {https://link.springer.com/article/10.1007/s12525-023-00636-5}
}

@article{rogalski2024worthIt,
  author  = {Rogalski, Timo and Schiereck, Dirk},
  title   = {When is blockchain worth it? Value and risk drivers of corporate blockchain announcements},
  journal = {Electronic Markets},
  year    = {2024},
  volume  = {34},
  number  = {1},
  pages   = {39},
  doi     = {10.1007/s12525-024-00718-y},
  url     = {https://link.springer.com/article/10.1007/s12525-024-00718-y}
}

\end{document}